\documentclass[fleqn]{aa}  
\usepackage{supertabular,booktabs}  
\usepackage{graphicx}
\usepackage{xcolor}
\usepackage{hyperref}
\usepackage{txfonts}
\usepackage{calligra}
\usepackage{calrsfs}
\usepackage[T1]{fontenc}
\usepackage[mathcal]{eucal}
\usepackage{longtable}
\usepackage{multicol}

\usepackage[switch, modulo]{lineno}

\sfcode`A=1000 \sfcode`B=1000 \sfcode`C=1000 \sfcode`D=1000
\sfcode`E=1000 \sfcode`F=1000 \sfcode`G=1000 \sfcode`H=1000
\sfcode`I=1000 \sfcode`J=1000 \sfcode`K=1000 \sfcode`L=1000
\sfcode`M=1000 \sfcode`N=1000 \sfcode`O=1000 \sfcode`P=1000
\sfcode`Q=1000 \sfcode`R=1000 \sfcode`S=1000 \sfcode`T=1000
\sfcode`U=1000 \sfcode`V=1000 \sfcode`W=1000 \sfcode`X=1000
\sfcode`Y=1000 \sfcode`Z=1000

\newcommand{\Msun}{\hbox{M$_\sun$}}

\newcommand{\wslap}{\texttt{WSLAP+}\,}

\begin{document}

   \title{JWST lens model for A370: A very low dark matter fraction for a brightest cluster galaxy and lensing properties for the Dragon arc.}
   \titlerunning{A new JWST-based lens model for The Dragon}
   \author{Jose M. Diego \inst{1}\fnmsep\thanks{jdiego@ifca.unican.es}
       \and Fengwu Sun \inst{2} 
       \and Jose M. Palencia \inst{1}   
       \and Xiaojing Lin \inst{3,4}
       \and Marceau Limousin \inst{5}  
       \and Rachel Gledhill \inst{6}
       \and Anna Niemiec \inst{7}
       \and Wenlei Chen \inst{8}
       \and Rogier A. Windhorst \inst{9}
       \and Mitchell F. Struble \inst{10}
       \and Tom Broadhurst \inst{11,12,13}
    }      
   \institute{Instituto de F\'isica de Cantabria (CSIC-UC). Avda. Los Castros s/n. 39005 Santander, Spain 
        \and Center for Astrophysics $|$ Harvard \& Smithsonian, 60 Garden St., Cambridge, MA 02138, USA   
        \and Department of Astronomy, Tsinghua University, Beijing 100084, China  
        \and Steward Observatory, University of Arizona, 933 N Cherry Avenue, Tucson, AZ 85721, USA  
         \and Aix Marseille Univ., CNRS, CNES, LAM, Marseille, France 
         \and Cosmic Dawn Center (DAWN), Denmark 
         \and Univ. Grenoble Alpes, CNRS, Grenoble INP*, LPSC-IN2P3, 38000 Grenoble, France 
         \and Department of Physics, Oklahoma State University, 145 Physical Sciences Building, Stillwater, OK 74078, USA 
         \and School of Earth and Space Exploration, Arizona State University, Tempe, AZ 85287-6004, USA 
         \and Department of Physics and Astronomy, University of Pennsylvania, 209 South 33rd Street, Philadelphia, PA 19104, USA 
         \and Department of Theoretical Physics, University of Basque Country UPV/EHU, Bilbao, Spain
        \and Ikerbasque, Basque Foundation for Science, Bilbao, Spain
        \and Donostia International Physics Center, Paseo Manuel de Lardizabal, 4, San Sebasti´an, 20018, Spain
          }
 \abstract{
 We present a new lens model for the $z=0.375$ galaxy cluster Abell 370 based on previously spectroscopically confirmed lensed galaxies and new lensed systems identified in JWST data, including recent data from the MAGNIF program. 
 Based on the best models able to reproduce two radial arcs near the BCGs, we compare the stellar mass to the total mass from the lens model and find that the fraction of dark matter in the south BCG is consistent with $\Lambda$CDM while in the north BCG we find a very small amount of dark matter, more consistent with alternative models to $\Lambda$CDM. We discuss possible causes for this and conclude that additional data is needed to clarify the situation. 
 We study the lensing properties, magnification, time delay and strength of the critical curve, along the Dragon arc, where previous studies have reported tens of alleged microlensing events from supergiant stars at $z=0.7251$. The new lens model is able to reproduce the distribution of microlensing events with great accuracy. Some of the microlensing events may be reinterpreted as long-period Cepheid in future observations. We consider this possibility and study in more detail the challenges for such detection from intervening microlenses.
   }

   \maketitle
%
\section{Introduction}

In recent years, the \textit{James Webb} space telescope (JWST) has been re-observing some of the best gravitational lenses previously studied with the \textit{Hubble} space telescope (HST). Of special relevance are the six clusters from the Hubble Frontier Fields (HFF) program \citep{Lotz2017} where images reaching AB $\sim 28.5$ were obtained in seven filters between $\approx 0.4\, \mu m$ to $\approx 1.6\, \mu m$. This high quality data allowed to identify hundreds of multiply lensed galaxies behind these six clusters. Many of the new lensed galaxies were spectroscopically confirmed from the ground, with the Multi Unit Spectroscopic Explorer \citep[MUSE, ][]{Bacon2010} playing a pivotal role. 

One of these lenses is the galaxy cluster Abell370 (or A370), at redshift $z=0.375$, and containing the first discovered giant arc \citep{Soucail1987,Paczynski1987}. Previously known as a powerful source for X-rays \citep{Henry1982}, this cluster was selected as one of the HFF clusters due to its large mass and existence of known gravitationally lensed galaxies \citep{Mahdavi2001,Kneib1993,Richard2010,Richard2014,Johnson2014}. As part of this program,  the central $\sim 10$ arcmin$^2$ region was observed in wavelengths ranging from 0.45 $\mu$m to 1.6 $\mu$m and to a depth of $\sim 28.5$ mag in the visible and IR bands. The area observed by the HFF program around A370 was later doubled thanks to the Beyond Ultra-deep Frontier Fields and Legacy Observations (BUFFALO) program \citep{Steinhardt2020}, although with shallower observations than in the HFF program.

The HFF data from A370 have been extensively used by lens modelers \citep{Lagattuta2017,Diego2018b,Lagattuta2019,Ghosh2021,Lagattuta2022,Li2024,Limousing2025}. BUFFALO data were used to extend the lens model to larger radii by including weak lensing measurements from HST \citep{Niemiec2023}. JWST has observed A370 as part of different programs (GTO-1208 or CANUCS, GO-2883 or MAGNIF, and GO-3538). Based on CANUCS data,  \cite{Gledhill2024} confirms a new lensed galaxy at $z\sim 8$ and refined a parametric lens model with all the available constraints. No further new lensed systems were found in the cluster field, "likely due to already very deep HST" according to the authors.

Earlier lens models find that A370 is a merging cluster with the two main clumps having comparable masses and with the mass distribution and X-ray emission from Chandra in fairly good agreement. Smaller groups surround the two main clumps to the north and east. At $z\approx 1$ there is an overdensity of galaxies \citep{Diego2018b,Lagattuta2019} that could be playing some lensing role in some of the high-redshift lensed galaxies. Near the peaks in the mass distribution, some of the parametric models as well as some free-form models, find odd distributions for the dark matter (DM) in this cluster, in particular near the north BCG where parametric models require the existence of a DM halo with a significant offset of $\approx 50$ kpc with respect to the northern BCG \cite{Lagattuta2019} or  $\approx 35$ kpc in the case of free-form models \citep{Ghosh2021}. More recently, \cite{Limousing2025} finds that a dark clump halo with significant mass between the two BCGs is needed in their parametric model. Previously, \citep{Diego2018b} had shown how the observed position and morphology  of two lensed images near the BCGs can be explained without offsets in the DM distribution, but with the BCGs containing a relatively small amount of DM, and with their mass being largely accounted for by just their baryonic content. Galaxies with small amounts of DM have been claimed in the literature, with their low DM content explained as being a relic (and originally gas rich) early type galaxy where accretion has stopped \citep{Comeron2023}, or the result of tidal interactions \citep{Ai2025}, or satellites falling into larger halos \citep[see for instance halo 6 in][]{Flores-Freitas2022}. However, no mechanism is known that could expel large amounts of DM  from the deep potentials at the center of galaxy cluster. Revisiting the DM content in the two BCGs of A370 is one focus of this paper.

\begin{figure*} 
  \includegraphics[width=\linewidth]{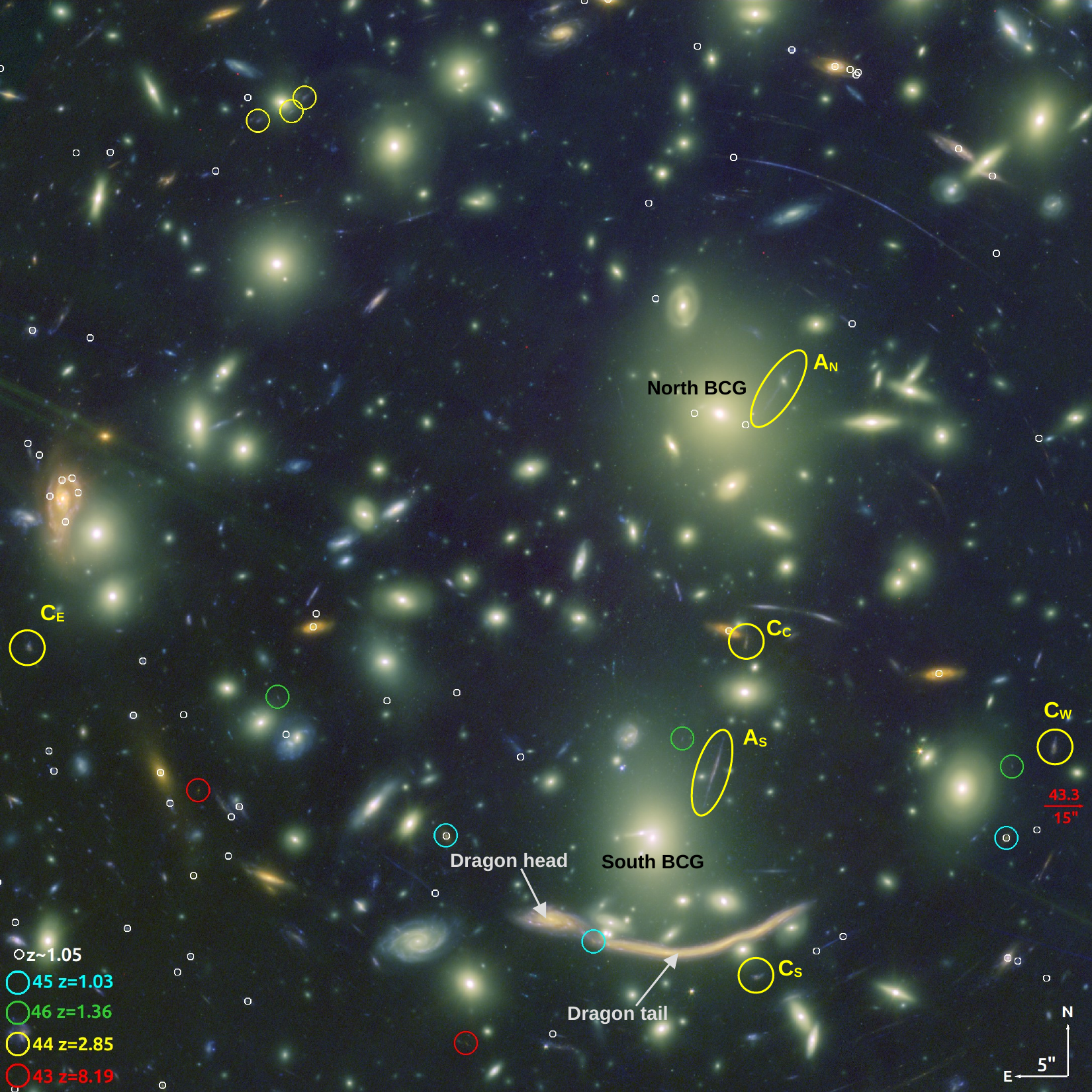} 
      \caption{Color image of A370 combining 16 HST and JWST filters (range 0.2--5 micron).  The scale and orientation are indicated. 
       Medium-size colored circles mark four new systems (or redefinitions) used in the lens model with redshifts obtained from MAGNIF. Counterimage 70.3 is outside this image $\approx 15$ arcseconds west. Small white circles mark galaxies with MAGNIF photo-z $1.05\pm0.05$ from a lensed group overdensity behind the cluster. Some members of this group are multiply lensed. The two BCGs and the Dragon galaxy are marked. The two big ellipses show the location of two radial arcs (A$_{\rm N}$ and A$_{\rm S}$) from system 7. Each radial arc contains a double image of different portions of the source galaxy in system 7. The yellow circles labeled C$_{\rm E}$,  C$_{\rm S}$,  C$_{\rm C}$, and  C$_{\rm W}$ contain additional counterimages of the source in system 7. The straight features in the NE are diffraction spikes from nearby bright stars.
         }
         \label{Fig1}
\end{figure*}

More recently, JWST has also observed A370 through multiple programs, most noticeably as part of the CAnadian NIRISS Unbiased Cluster Survey (CANUCS; GTO-1208, PI: Willott) in Cycle-1, Medium-band Astrophysics with the Grism of NIRCam in Frontier Fields (MAGNIF; GO-2883, PI: Sun) and GO-3538 (PI: Iani) in Cycle-2.
These observations extended the wavelength coverage to $\lambda \approx 5$\,$\mu$m, and with a superior resolution and sensitivity in the IR range than previously observed by HST.
A color-image of the central region of A370 combining data from JWST with previous data from HST is shown in Fig.~\ref{Fig1}. This color composite combines the information from 16 filters with spectral information ranging from $\approx\,0.2\,\mu$m to $\approx\,5\, \mu$m.

One of the more surprising results obtained with the JWST data on A370 was the discovery of a record-breaking number of microlensing events. A370 hosts a remarkable lensed galaxy nicknamed The Dragon, at $z=0.7251$. The relatively low redshift of the Dragon combined with the fact that the lensed galaxy is relatively large and intersects several caustics in the source plane results in an unprecedented large number of transients. Most of them are believed to be supergiant (SG) stars in the Dragon undergoing microlensing events. In \cite{Fudamoto25} the authors report more than 40 such events from comparing two epochs of JWST data. Without lensing, SG stars in these galaxies would have apparent magnitudes in the range 35--38, and out of reach of JWST. During microlensing events in this arc magnification factors can reach between $10^3$ and $10^4$ (equivalent to a 7.5--10 mags increase) and some of these SG stars could be detected by JWST. Multi-epoch deep observations of this arc carried out with JWST will reveal many more events, including possibly the first Cepheids at cosmological distances and even reach fainter stars in the tip of the red giant branch \citep{Diego2025}. Most of the transients in \cite{Fudamoto25} are found close to the critical curve (CC) of the cluster, where the temporary alignment between stars responsible for the intracluster light (ICL), and stars in the background galaxy, increase the flux of the background stars by a few magnitudes thanks to microlensing.

In this paper we revisit this cluster presenting a new hybrid lens model derived using the most updated constraints of A370. Based on the JWST NIRCam data processed by the MAGNIF team, we identify several new candidates and estimate their redshift from the lens model, which is later compared to the photometric redshift derived with all the available photometric information (up to 19 filters from HST and JWST, 16 of them in the cluster core). The paper focuses on two aspects of the lens model: i) the DM content of the two BCGs and ii) the lensing properties along the Dragon arc.
The paper is organized as follows.
Section~\ref{s:MAGNIF} describes the JWST observations and data processing. 
This section includes also a discussion of the lensing constraints (including the new ones discovered by JWST) used to derive the lens model.
Section~\ref{s:WSLAPplus} describes very briefly the algorithm \wslap used to derive the lens model.
In section~\ref{sect_BCGs} we focus our attention on the two radial arcs near the south and north BCG and compare the stellar mass with the DM mass. Section~\ref{sect_Dragon} pays attention to the lensing properties of the Dragon arc. 
We discuss our results in section~\ref{sect_discussion}, in particular the implications for the DM content in the two BCG for MOND models and the possibility (and challenges) of detecting a particular type of variable object that could be interpreted as a transient, Cepheids, in the Dragon arc.  Finally, Section~\ref{sect_conclusions} summarizes our conclusions.

We adopt a standard flat cosmological model with $\Omega_M=0.3$ and $h=0.7$. At the redshift of the lens ($z=0.375$), and for this cosmological model, one arcsecond corresponds to 5.162~kpc. For the redshift of the Dragon ($z=0.7251$) one arcsecond corresponds to 7.245~kpc and the distance modulus is 43.2. 
Unless otherwise noted, magnitudes are given in the AB system \citep{Oke1983}. Common definitions used throughout the paper are the following: ``macromodel''  refers to the global lens model derived in Section~\ref{s:WSLAPplus}. The main caustic and main CCs are from this model. We use the term ``micromodel'' to refer to perturbations introduced by microlenses in the lens plane (stars and remnants) and  the term ``microcaustics'' to refer to regions in the source plane of divergent magnification associated with microlenses.

\section{JWST data and lensing constraints}
\label{s:MAGNIF}
A370 was observed in three different epochs by JWST from December 2022 to August 2024 as part of three programs (CANUCS, GO-3538, and MAGNIF).
NIRCam imaging observations were obtained in 15 filters, including eight filters from Cycle-1 CANUCS (F090W, F115W, F150W, F200W, F277W, F356W, F410M, F444W) and additional seven medium-band filters by the two Cycle-2 programs (F182M, F210M, F300M, F335M, F360M, F460M, F480M).
NIRCam imaging data reduction was performed by the MAGNIF team (Sun, F.\ et al., in prep.) through a modified stage-1/2/3 JWST pipeline \citep{bushouse_2023_7829329} \verb|v1.11.2| and CRDS calibration reference \verb|jwst_1188.pmap|.
We stacked the NIRCam data from F182M to F444W band to produce a detection image, extracted and deblended sources from the detection image and obtained aperture-photometry catalog of sources in the A370 field.
Photometric redshifts of galaxies were estimated using \textsc{eazy} \citep{Brammer2008} using $r=0.1^{\prime\prime}$ photometry and the templates from \citet{hainline24}. 
To correct for contributions from the ICL, we subtract the local background using a rectangular annulus (of width 5 pixel or $0".15$) outside the Kron aperture. Detailed NIRCam data processing has also been described by \citet{FuS2025}.

This work focuses only on the cluster core field which is the only one containing lensing constraints. A portion of the area observed by JWST in this cluster core region is shown in Fig.~\ref{Fig1}. This image (and others in this paper) is a color composition made after combining previous HST data from the HFF and FLASHLIGHTS programs with the new JWST data. In particular, the blue channel contains HST's F200LP, F435W, F606W, and F814W, the green channel is a combination of JWST's F090W, F115W, F150W, F182M, F200W, and F210M filters, and the red channel includes F277W, F300M, F335M, F356W, F460M, and F480M. Between HST's F200LP and JWST's F480M there is a factor $\approx 40$ in wavelength. Computing the effective depth of the stacked images is beyond the scope of this paper, but since some of the filters in each of the red, green and blue channels reach AB$\sim29$, it is safe to assume that the stacked images are deeper than AB 30. 
In particular, the depths of each individual filter are 29.4, 29.3 and 29.7 in HST's F435W, F606W, and F814W respectively, while for the JWST filters they are 28.7, 28.8, 29.0, 28.7, 28.6, 28.4, 29.4, 29.4, 28.8, 28.5, 29.0, and 27.4 for F090W, F115W, F150W, F182M, F200W, F210M, F277W, F356W, F360M, F410M, F444W, ands F480M, respectively. These correspond to the median $5-\sigma$ point-source depths measured with an aperture of diameter ${\rm D}=0".3$ in a region near the cluster center ($0.5'<r<1.5')$. These depths are comparable to the ones reported by the CANUCS collaboration in \cite{Sarrouh2025}\footnote{see their Table 2 listing $3-\sigma$ point-source depths to be compared with our $5-\sigma$ depths}.

\begin{table}
\caption{Redefinition of previous lensed families (11, 21 and 43) and new multiple images (44, 45, and 46). \label{tab_arcs}}
\label{table:1}      
\centering                                      
\begin{tabular}{c c c c}          
\hline\hline                        
     ID  &     RA    &    DEC     &    $z_{\rm MAGNIF}$   \\  
\hline                                   
   11.3 & 39.9884408 &  -1.5719997  &  7.58      \\ 
   21.3 & 39.9823189 &  -1.5815119  &   --       \\ 
   43.3 & 39.9564600 &  -1.5816400  &  8.12      \\ 
   44.1 & 39.9798079 &  -1.5642217  & 2.82       \\ 
   44.2 & 39.9801208 &  -1.5645528  & --         \\ 
   44.3 & 39.9809379 &  -1.5647886  & 2.97       \\ 
   45.1 & 39.9763554 &  -1.5822175  & 1.03       \\ 
   45.2 & 39.9626800 &  -1.5822717  & 1.1        \\ 
   45.3 & 39.9727517 &  -1.5848044  &  --        \\ 
   46.1 & 39.9705883 &  -1.5798517  &  1.36      \\ 
   46.2 & 39.9625450 &  -1.5805317  &  1.61      \\ 
   46.3 & 39.9804654 &  -1.5788375  &  1.44      \\ 
\hline                                             
\end{tabular}
\end{table}

For the lensing constraints we use the latest compilation from \cite{Gledhill2024} which is almost fully derived from previous HST observations plus spectroscopic confirmations, mostly by MUSE. The compilation in \cite{Gledhill2024} contains 43 multiple images. The set of lensing constraints is almost identical to the constraints used in \cite{Niemiec2023}.
We visually inspect all the constraint positions in our deep color images (HST+JWST) and reject those that do not show up as a source in the HST+JWST images. These are mostly MUSE detections with no counterpart in our deep images or in portions of the lens plane where contamination by members (for instance near the BCGs) makes it impossible to see the alleged lensed galaxy. These are systems 15, 30, 31, 32, 33, 34, 35, 36, 37, 41, and 42. None of these systems is ranked as gold in \cite{Gledhill2024} and hence not used in that work to derive the lens model. 
The resulting set of constraints represents our gold sample and is used to derive the lens model. In addition to these, we update the high-z system 43 with a third counterimage (43.3) not identified in earlier work (outside of the CANUCS FoV) but robustly confirmed with the MAGNIF data (both via photometry and morphology) as well as by its geometric redshift. We also update 11.3 with a new counterimage that matches better the colors of 11.1 and 11.2. We also update the coordinate of 21.3 based on the new MAGNIF reduction. In addition to these, we include two additional lensed galaxies (new systems 45 and 46) near the Dragon arc and the south BCG. Although these two galaxies are not confirmed spectroscopically, they are robust in terms of morphology and photometric redshift. These two galaxies are relevant for modeling the lens near the Dragon arc and are highly consistent with a preliminary lens model that does not include these galaxies as constraints.  Finally, we also include the triply lensed supernova (SN) in \cite{Chen2022} as the new system 44 that was not considered as a constraint in \cite{Gledhill2024}. 
The new redefinition of system 43 at $z=8.19$, the triply lensed SN at $z=2.85$ and the two new additional families of galaxies are marked with circles in Fig.~\ref{Fig1}. The lensed family at $z\approx 1$ belongs to a $z\sim1$ overdensity that was previously identified using HST  \cite{Diego2018b} and MUSE \cite{Lagattuta2019} data. JWST data confirms the existence of such overdensity behind the cluster (and magnified by it) and identifies new members. One of the galaxies in this group is multiply lensed into three counterimages, with one of them overlapping the Dragon and hence confirming this galaxy is behind the Dragon and hence not playing any role in the lens model for the Dragon arc.  The coordinates of these additional constraints are listed in Table~1.  \\

\section{Lens models}\label{s:WSLAPplus}
Using the constraints described in the previous section, we derive the lens model with the code \wslap \citep{Diego2005,Diego2007,Sendra2014,Diego2016}. This is a hybrid type of model that combines a free-form decomposition of the smooth, large-scale component (DM, gas, and lower-density stellar contribution to the intracluster light) with a small-scale contribution from cluster galaxies (higher density stellar component in member galaxies). The code combines weak (when available) and strong lensing in a natural way with the large-scale and small-scale components in the mass equally affecting the strong- and weak-lensing observables. For this work, only strong-lensing constraints are used.

The algorithm assumes the smooth mass is described by a superposition of Gaussian functions distributed over a predetermined grid, while the compact mass component follows the distribution of light around member galaxies which are identified from the spectroscopic sample of \cite{Lagattuta2019}. The members can be distributed in different groups or layers, each one assuming a constant mass-to-light ratio. We derive several models by making small adjustments to the variables in the optimization. In particular we are interested in exploring the range of solutions that best describe the two radial arcs from system 7. The two radial arcs from system 7 are the most sensitive to the distribution of mass near the two BCGs.  We use these arcs to define a model taxonomy based on how well a model is able to reproduce these arcs. We produce three lens models that we refer to as A, B, and C. All three models have the same grid configuration for the distribution of Gaussians. The grid for the Gaussians is derived after iterating the solution three times, where in the first iteration we start with a regular grid of $25\times25$ identical Gaussians. In a second iteration we build a new grid increassing the resolution (that is, increasing the number of Gaussians and reducing their width) in the portions of the lens that contain more mass. The third and final iteration repeats the process with the solution obtained in the second iteration. This final grid has a total of 308 grid points where we place Gaussians with varying widths.  The smallest Gaussians have a width of $\approx 10"$ and are found near the two BCGs, while the largest Gaussians have widths $\approx 60"$ and are found in the corners of the modeled area ($3'.2\times3'.2$).

For model A, all member galaxies are placed in one single layer, that is, the mass-to-light ratio is the same for all galaxies in A370. In models B and C we allow the two BCGs to have their own mass-to-light ratio. Model B has the north BCG in layer 1, the south BCG in layer 2, and the remaining member galaxies in layer 3. Model C is similar to model B (three layers) with the only difference that we add two more Gaussians with widths $4".36$  at the position of the two BCGs in order to add more flexibility in the lens model at the BCG positions. 
Since the constraints are the same and the differences in configuration are relatively small,  models A, B, and C are very similar but show some differences around the two BCGs which is the region we are interested in exploring in more detail (see next section). The RMS for models A, B, and C are 1", 0".96, and 0".98 respectively. For comparison, \cite{Niemiec2023} quotes an RMS of 0".9 for their parametric model and the best model in \cite{Limousing2025} has an RMS of 0".7. This model has similar lensing constraints to our model and the model in \cite{Niemiec2023}. In addition to the A, B, and C models, we produce an additional fourth model, D, that is optimized for the Dragon arc. This model uses as additional constraints the location of critical points in the Dragon arc as derived from visual inspection of the arc. We discuss this model in detail in section~\ref{sect_Dragon}.

\begin{figure} 
  \includegraphics[width=9cm]{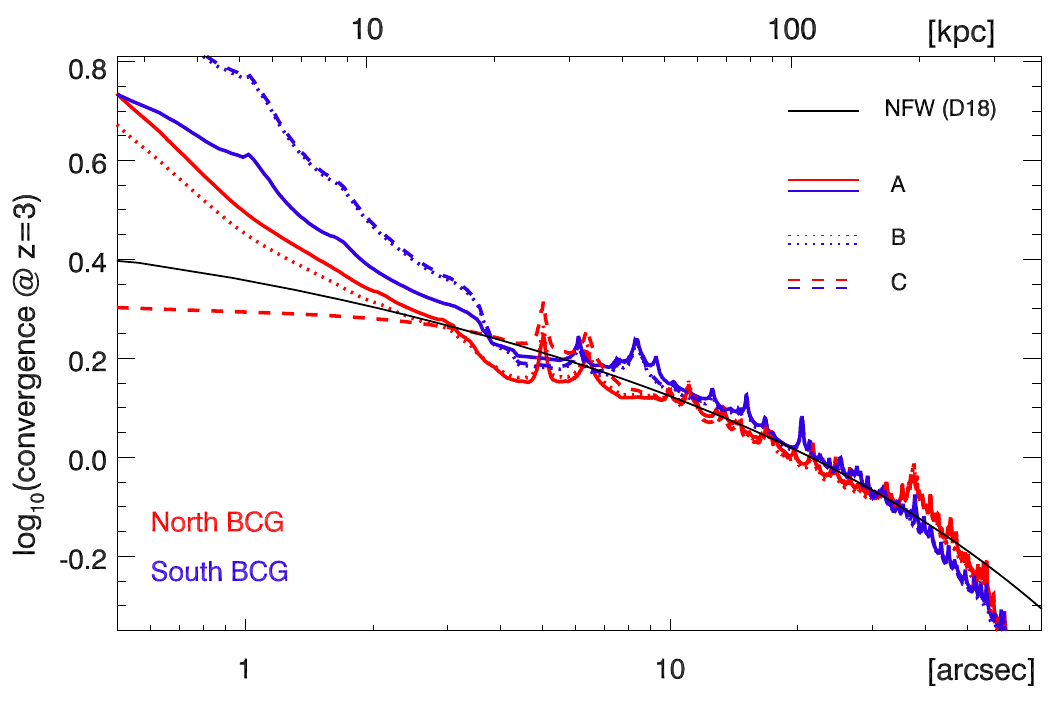}   
      \caption{Mass profile of the lens models A, B, and C. 
       Red curves correspond to the radial profiles when the center is the north BCG while blue curves are for the profiles centered in the south BCG. The black lines is the same NFW model from \cite{Diego2018b} with concentration $c=2$ and a virial radius $R_{200}= 3$ Mpc. 
       The profiles around the north and south BCG are remarkably similar, suggesting a mass ratio of 1:1 for the two subgroups.
         }
         \label{Fig_Profile}
\end{figure}

\section{The dark matter content of the BCGs}\label{sect_BCGs}
The distribution of mass for models A, B, and C are all very similar with the main differences concentrating (unsurprisingly) around the two BCGs, as expected given the fact that the variables in the three models were picked to explore the uncertainty in the lens model around the BCGs, and the number density of constraints near the BCGs is smaller than in other areas of the cluster. The profiles centered in each of the BCGs for the models A, B, and C are shown in Fig.~\ref{Fig_Profile}. 
The black solid line is the same NFW model in \cite{Diego2018b} showing consistency with earlier models. In general all models agree well in terms of mass profile above $\approx 20$ kpc from the center. The lens model shows also a very similar distribution of mass around the north and south BCG, confirming earlier results relative to the similar mass of the two subgroups. 
 
To better assess the quality of the different lens models near the BCG, we focus our attention on the two radial arcs closest to the two BCGs. Both radial arcs belong to the same family of lensed galaxies (system 7 at $z=2.7512$) as first suggested by \cite{Diego2018b} and later confirmed spectroscopically by MUSE \citep{Lagattuta2019}. Interestingly, these two radial arcs are not pointing straight to the BCGs as one would expect if the BCGs were very massive. Instead, they are aligned in directions which miss the BCG centers by 3".1 and 3".6 for the north and south BCGs respectively, thus naively suggesting that the BCGs may have relatively shallow potentials. Previously, \citep{Diego2018b} had shown how the observed position and morphology  of the two radial arcs can be better explained if the BCGs contain only a small amount of DM, and with their mass being largely accounted for by their baryonic content. 
Some of the models in Fig.~\ref{Fig_Profile} are shallow so we can check if in fact these models are able to better reproduce the arcs.

\begin{figure} 
  \includegraphics[width=\linewidth]{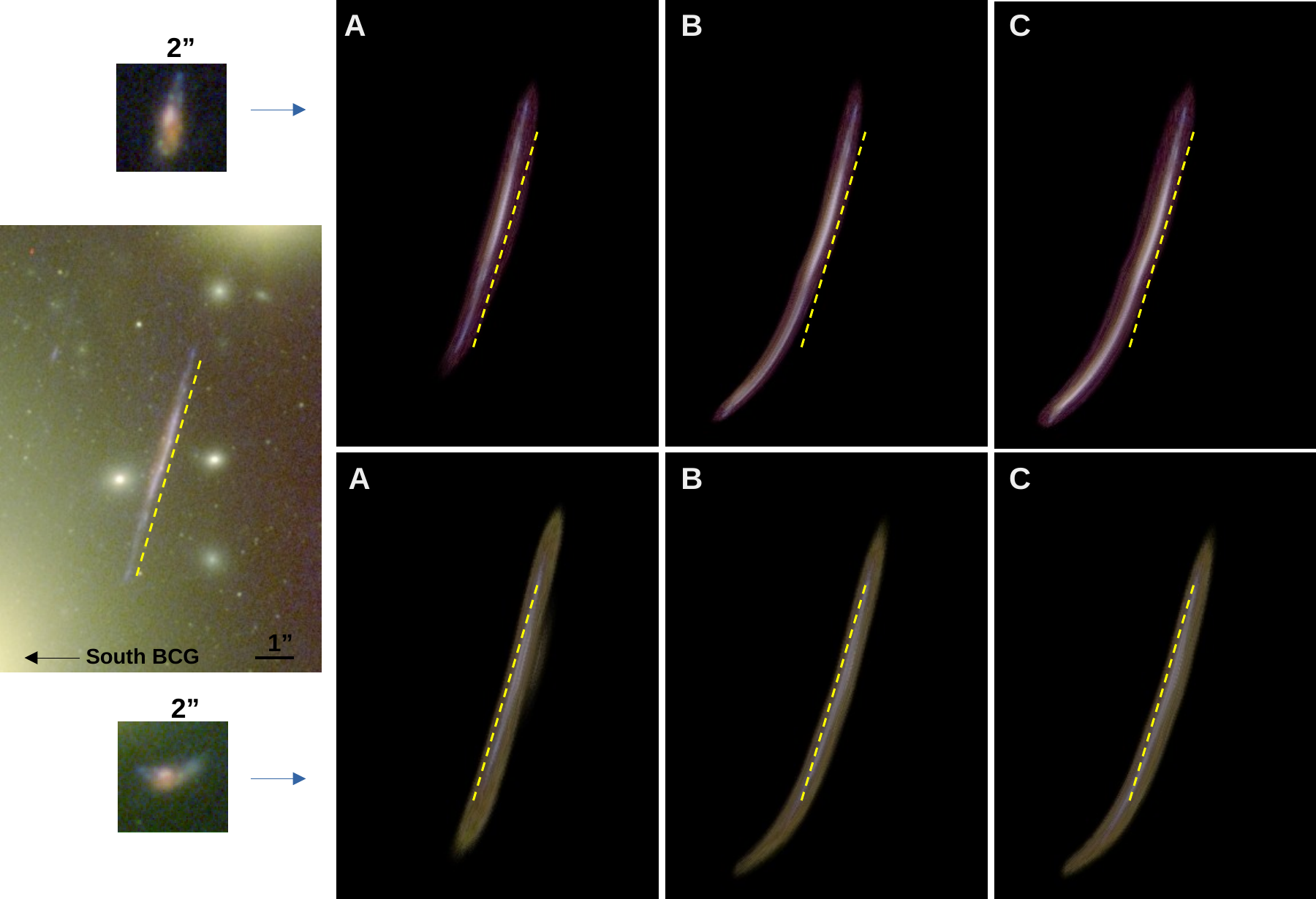}    
      \caption{Predicted radial arc 7a near the South BCG for models A, B, and C, and for two different counterimages used as templates.  The dashed yellow line is identical in all panels and can be used to compare with the position and orientation of the observed arc.
         }
         \label{Fig_System7a}
\end{figure}

In Fig.~\ref{Fig_System7a} we compare the observed radial arc near the south BCG with the prediction by the three lens models in Fig.~\ref{Fig_Profile}. For each model we use two different counterimages (the least distorted and closest to the radial arcs) to derive the template of the galaxy in the source plane. One of the counterimage templates is on the west side of the cluster and the other one is on the south. A slightly distorted third counterimage in the east side is not used because it is much farther away from the radial arcs, and hence more sensitive to portions of the lens plane that are not so well constrained. Each template is delensed into the source plane and then lensed back into the image plane. The top row in Fig.~\ref{Fig_System7a} shows the predicted radial arc when the counterimage on the west side of the cluster (shown in the small square in the top left) is used as template. The bottom row is the same but using the counterimage in the south (shown in the small square in the bottom-left) as template. In all panels we include a dashed yellow line with the location and orientation of the observed radial arc. All panels and the dashed line have the same size and are centered in the exact same location as the panel on the left (the observed arc). From the figure we see in both cases (west and south templates) that models B and C predict the bottom of the radial arc to curve toward the BCG. This curvature is not observed in the data so we can argue that models B and C contain too much mass around the south BCG and are not consistent with the observations. 
From the comparison with the observed south radial arc, we can conclude that for the south BCG, model A makes the best prediction for the radial arc near this BCG. This model (solid dark-blue in Fig.~\ref{Fig_Profile}) has the shallowest profile, confirming that the data prefers a model with less mass for the south BCG.

\begin{figure} 
  \includegraphics[width=\linewidth]{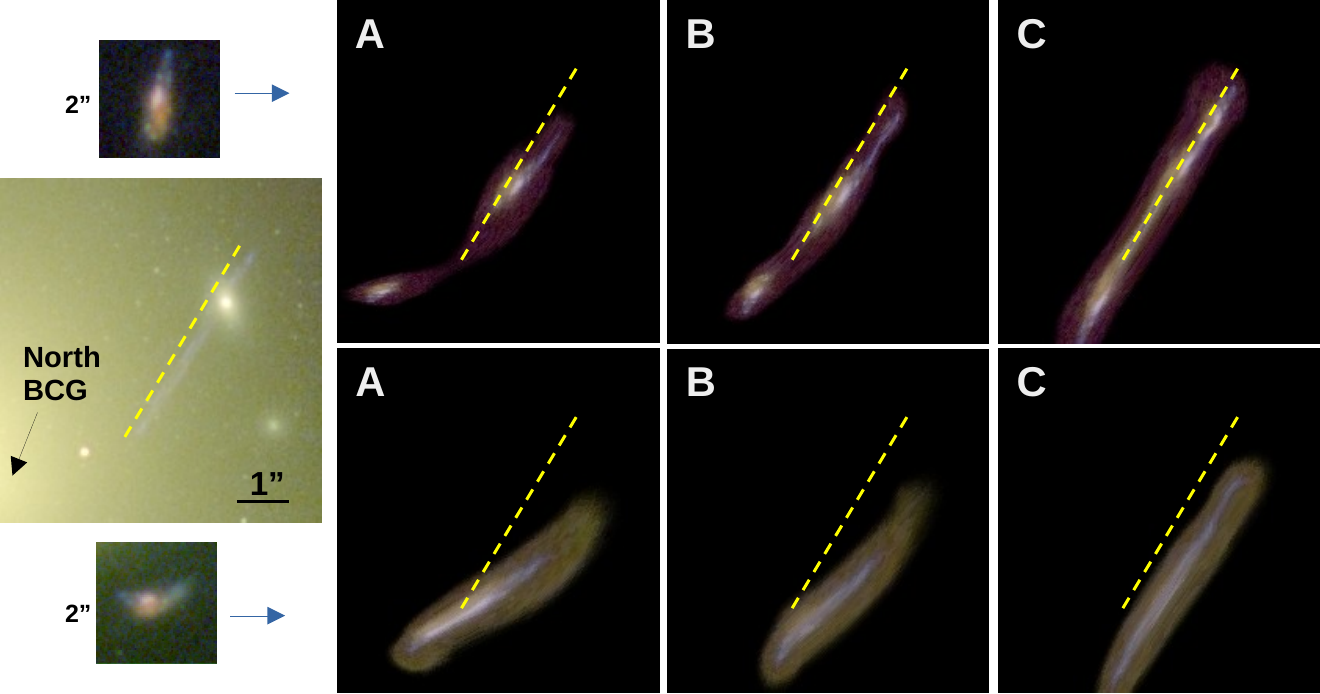}    
      \caption{Predicted radial arc 7b near the North BCG for models A, B, and C, and for two different counterimages used as templates.  The dashed yellow line is identical in all panels and can be used to compare with the position and orientation of the observed arc.
         }
         \label{Fig_System7b}
\end{figure}

In Fig.~\ref{Fig_System7b} we repeat the same exercise but for the radial arc near the north  BCG. In this case all models except model C fail at reproducing the morphology of the arc. For model C, when the counterimage on the west is used as template, the radial arc appears in the right position and with the right orientation. However, the predicted image is not perfect as the south and central portions of the predicted arc do not appear in the observed image. The observed arc only shows the northern blue tip of the galaxy imaged into a merging double image, while the predicted image shows a larger portion of the galaxy. When we use the south counterimage as template, this problem disappears since only the tip of the template image is multiply lensed, although in this case the arc is predicted to form $\approx 1"$ south of the observed position. Despite this small offset, model C provides a reasonably good fit to the arc once we consider the two counterimage templates. 
From Fig.~\ref{Fig_Profile}, the profile for model C at the north  BCG (short-dashed red line) is very shallow, suggesting that the north BCG contains a relatively smaller amount of dark matter, but we recall that model C places a small Gaussian at the same position of the BCG. In the same figure we see how between 20 and 30 kpc, model C (red dashed line) contains more mass than the  other two models, A and B (red solid and dotted lines respectively), thus partially compensating the inner shallow profile. 
From this point forward, we consider only model A for the south BCG and model C for the north BCG. From these lens models we can estimate the mass associated to the north and south BCGs. 
From model A the total projected mass in a 23 kpc aperture around the south BCGs is 
$M_{\rm south}=6.001\times 10^{12}$ \Msun. Similarly, from model C the total projected mass in a 23 kpc aperture around the north BCGs is 
$M_{\rm north}=4.505\times 10^{12}$ \Msun.

Since some of the mass in the north and south BCG must be attributed to the baryonic component, in order to estimate the amount of dark matter contained within the two BCGs we need to compare it with the stellar mass. Based on HST data, \cite{Diego2018b} estimated a combined mass for the two BCGs of $1.15\times10^{12}$ \Msun. If one assumes they both have the same light-to-mass ratio, and since they have similar magnitude, the stellar mass of the south and north BCG should be approximately half the combined stellar mass, that is $M_*\approx 5.75\times10^{11}$ \Msun. We have revised these masses combining HST with the new JWST photometry. 
Our new estimate of the stellar mass are derived with {\small CIGALE}  \citep{Boquien2019}. We assume the Bruzual \& Charlot stellar population with metallicity in the range 0.2--2.0 Solar \citep{Bruzual2003} with a delayed-$\tau$ star formation history. The initial mass function is a Chabrier model \citep{Chabrier2003}. 
For the exact aperture we adopted in the compact component of the lens model ($\approx 23$ kpc for the two BCGs) we derive stellar masses of $M_*= (4.89\pm0.99)\times10^{11}$\Msun\, and $M_*= (4.36\pm0.76)\times10^{11}$\Msun\, for the north and south BCG respectively. These are $\approx 25\%$ smaller than our old estimate based on HST data, mostly due to the fact that the aperture considered for the BCG in the compact component is smaller (by $\approx 17\%$ in radius) in this work than in \cite{Diego2018b}. For the remaining of the discussion we adopt our new estimates for the stellar mass of the compact component  associated to the two BCGs. By dividing the total mass from the lens model by the stellar mass obtained in the same aperture we derive the total to stellar mass ratios.

These masses above are however projected along the line of sight. We are more interested in the volumetric mass within a radius 23 kpc (in 3D) but after correction for projection effects, that is after subtracting an estimate of the mass along the line of sight beyond 23 kpc from the center of the BCGs. We do this by applying a correction factor to both, the stellar mass and the total mass. For the stellar mass, first we estimate the average flux, $f_{23}$ in F200W (a different JWST filter showing prominent emission from the BCGs, such as F277W, gives similar results), at exactly 23 kpc distance, and compute the value, $\aleph_{23}={\rm N}\times f_{23}$, where ${\rm N}$ is the number of pixels within our aperture of 23 kpc. The value of $\aleph_{23}$ is an estimation of the flux (proportional to the stellar mass) we have along the line of sight just outside the considered aperture. Then we compute the total projected flux within that aperture, $\mathcal{T}_{23}=\sum_1^{\rm N}f_i$, where $f_i$ is the observed flux in pixel $i$ within our aperture. The correction factor is then $\mathcal{C}=(\mathcal{T}_{23}-\aleph_{23})/\mathcal{T}_{23}$. For the north BCG we find the stellar mass needs to be corrected by a factor $\mathcal{C}=0.663$ while for the south BCG the correction factor is $\mathcal{C}=0.708$. A similar calculation is made for the total mass but substituting the flux by the mass in the lens models (A for the south BCG and C for the north BCG). The total mass within 23 kpc for the north BCG (model C) needs to be corrected by a factor $\mathcal{C}=0.073$ while for the south BCG (model A) this is $\mathcal{C}=0.341$. The much smaller correction factor for the north BCG is due to the shallowness of the profile in model C, that places most of the mass along the line of sight outside the 23 kpc radius. Taking both correction factors into account we finally find that for the north BCG the total to stellar mass ratio is $(0.073\times4.505\times10^{12})/(0.663\times4.89\times10^{11})=1.014$ and for the south BCG we find a total to stellar mass ratio of  $(0.341\times6.001\times10^{12})/(0.708\times4.36\times10^{11})=6.62$. If we ignore the contribution to the mass from gas, which should be small within the 23 kpc, these numbers translate into DM fractions of $f_{\rm DM} \approx 0$ for the north BCG and $f_{\rm DM} \approx 0.85$ for the south BCG.

For the south BCG the result is higher than expectations in $\Lambda$CDM.  Using the relation from \cite{Santucci2022}, $f_{DM}\sim0.1+0.17\times({\rm log}(M_*/M_{\odot})-10.59)^2$, this relation predicts $f_{DM}\sim0.24$ for a stellar mass of $0.708*4.36\times10^{11}$ \Msun. The relation from \cite{Santucci2022} is for $R_{\rm h}=1$ while our aperture is closer to $2 R_{\rm h}$. Since  $f_{DM}$ grows rapidly with radius \citep[see for instance Fig.19 in][]{Comeron2023}, we expect $f_{DM}$ to grow above 0.24 although it is unclear by how much. Our derived value of $f_{\rm DM} \approx 0.85$ for the south BCG may be overestimated, for instance if the true profile is shallower than the one in model A, the correction factor $\mathcal{C}$  would be smaller, hence reducing also the value of $f_{\rm DM}$. 

For the north BCG the result is most surprising as it favors a galaxy with little to no amount of DM. This is similar to the conclusion in \cite{Diego2018b}. Considering model B, our second best model in terms of reproducing the north radial arc, repeating all the calculations above we find a total to stellar mass ratio of 5.6 for the north BCG and $f_{DM}=0.82$ for model B, very similar to the result for the south BCG in model A (the best model for the south radial arc). Recent lens models from \cite{Niemiec2023,Gledhill2024,Limousing2025} contain also high $f_{\rm DM}$ values for the north BCG, but as we show in the appendix, these models do not reproduce the radial arc near the north BCG in a satisfactory way, making model C the best in terms of reproducing this arc. We can also make the argument that if the true profile is steeper than model C, the correction factor discussed previously would be larger, increasing the value of $f_{\rm DM}$. 

Our results above are inconclusive but suggestive of interesting lessons to be learned from the two BCGs in A370. The DM content of the BCGs remains an open question, with the apparent dichotomy in our results relative to the BCGs only to be resolved with better modeling of the lensing effect or with additional data in or around the BCGs. For instance by identifying more counterimages near the two BCGs or by incorporating kinematic constraints from high resolution spectral observations around the BCGs. To test the possibility of a better model for the north radial arc, we derive a new model, C', that forces the compact component of the two BCGs to have exactly the estimated stellar mass (i.e., no longer a free parameter in the model). This is a priori a more physically motivated model than model C on its ontological sense (after all the BCGs must contain at least the stellar mass). We build this model by first subtracting the deflection field due to the stellar mass of the two BCGs from the observed arc positions, and then proceeding with the optimization using the same setting as model C, but with the only difference that the compact component no longer contains the two BCGs. For the south BCG, models C and C' are very similar with the smooth component peaking at the position of the BCG in both cases. Because of this similarity between C and C', we focus only on the northern BCG where the differences between models C and C' are more pronounced. 
In model C' the small Gaussian placed at the north BCG position contains a very small mass, orders of magnitude smaller than the stellar mass of the north BCG and consistent with zero. This is similar to the model C where only the stellar mass is needed in the inner 23 kpc.  The smooth component shows no local peak near, or at the position of, the north BCG. For this model we observe a poor reconstruction of the two radial arcs, similar to the results from \cite{Gledhill2024} (see appendix). Hence this model is not a good solution for the north radial arc and we consider model C as our best model, where the tails in the mass distribution of the small central Gaussian (at the BCG position) seem to be helping at reproducing the radial arc. The lessons from model C' are valuable since the mass of the small Gaussian at the north BCG is incredibly small. The algorithm is trying to push this mass (the only free parameter at that location) as small as possible and would be negative if it was not for the constraint in \wslap that all masses need to be positive. The worse performance of model C' when compared with model C and the fact that both models have approximately the same mass within the considered aperture of 23 kpc suggests that the difference must be in the way this mass is distributed (or profile) beyond the 23 kpc radius,  with the shallower profile of model C being able to reproduce better the northern radial arc. Model C also shows a peak in the smooth distribution that is offset from the BCG. This peak is not present in model C' and may play a role in describing the northern radial arc. We discuss this offset in more detail in section~\ref{sect_discussion}.

\begin{figure*} 
  \includegraphics[width=\linewidth]{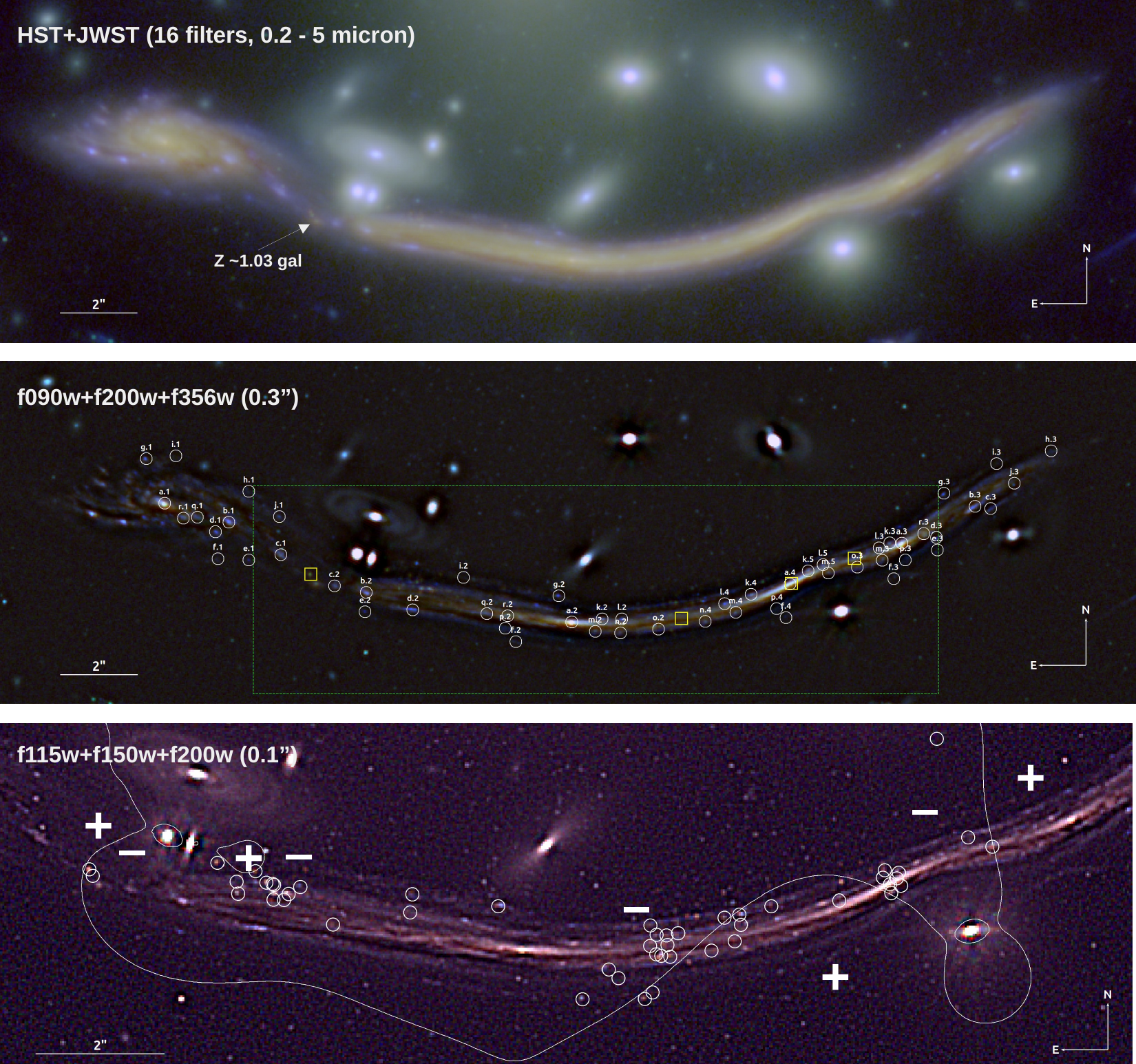}    
      \caption{Top. Composite HST+JWST image showing the Dragon arc region. The image is a made form 16 filters with the blue containing HST's F200LP, F435W, F606W and F814W, green with JWST's F090W, F115W, F150W, F182M, F200W, and F210M, and red with F277W, F300M, F335M, F356W, F460M, and F480M. Middle,  high-pass filtered color version (scale of 0.3 arcseconds for the high pass Gaussian filter) of the three filters indicated in the top-left corner. The circles mark the constraints in the Dragon used to derive the lens model. Yellow squares mark the position of the critical points also used as constraints. Bottom, high pass filtered (scale 0.1 arcseconds for the high pass Gaussian filter) color version of the three bands indicated in the top-left corner. The portion of the Dragon shown in this image is marked with a rectangle in the middle panel. Circles indicate the position of the transients in \cite{Fudamoto25}. The white curve is the hybrid \wslap lens model derived with critical points as constraints. The "+" and "-" symbols indicate the parity of the counterimages formed on that side of the CC. 
         }
         \label{Fig_Dragon}
\end{figure*}

\section{Lensing effects in the Dragon arc}\label{sect_Dragon}

In addition to the A, B, and C lens models discussed in section~\ref{s:WSLAPplus}, we derive a fourth lens model that is better suited for the Dragon. None of the three models discussed in section~\ref{s:WSLAPplus} produces a satisfactory CC around the Dragon. These models predict the CC crossing the Dragon arc  through points which are not symmetry points (the CC from these models are very similar to the model from \cite{Gledhill2024} and some of the models in \cite{Limousing2025} shown in the appendix). To remedy this, we derive a fourth model that includes additional constraints along the Dragon arc, including the position of critical points. The latter can be added as linear constraints after a suitable rotation of the deflection field by an angle given by the orientation of the arc \citep[see ][ for details on how to add critical points to the set of constraints]{Diego2022_Godzilla}. The full set of constraints in the Dragon, including the location of the critical points, is shown in the middle panel of Fig.~\ref{Fig_Dragon}. Naively, one would think this solution is superior to the other solutions presented in section~\ref{s:WSLAPplus} but this is only the case around the region of the Dragon arc. A solution derived with just critical points (in the Dragon and in other arcs where critical points can be identified based on symmetry arguments) is able to produce lens models with critical curves passing through the critical points but the mass distribution does not resemble the mass distribution of models A, B, C. The mass obtained with just critical points predicts clumps of mass at locations that do not correlate well with the location of prominent member galaxies, hence making the solutions based on only critical point information unreliable. A solution obtained with both, the location of multiply imaged galaxies and critical points resembles a combination of the solutions obtained with either set of constraints, and is in general better than the models A, B, and C in some portions of the lens plane, and worse in other portions. Some of the artifacts from the critical-point only solution appear also in the solution obtained with both constraints. 
In situations where the cluster is relatively symmetric, as it is the case of the cluster lensing the Sunburst arc, the addition of critical points along the Sunburst arc does always improve the solution, but in irregular clusters such as A370, this is not always the case and critical point information needs to be used with care and only locally, as we do here for the Dragon arc. The critical points in Fig.~\ref{Fig_Dragon} (yellow squares in the middle panel) correspond to symmetry points identified in the JWST images. The observed orientation of the arc used to rotate the deflection field is also derived from the JWST images.

The CC from the solution obtained after adding critical points in the Dragon is shown in the bottom panel of Fig.~\ref{Fig_Dragon} (white line). 
The circles in the figure show the location of the 40+ transients in \cite{Fudamoto25}. Our fourth model  places the CC very close to the position of these transients, thus reinforcing the interpretation that they are indeed microlensing events, which are expected to take place much more likely near CCs.

\begin{figure} 
  \includegraphics[width=\linewidth]{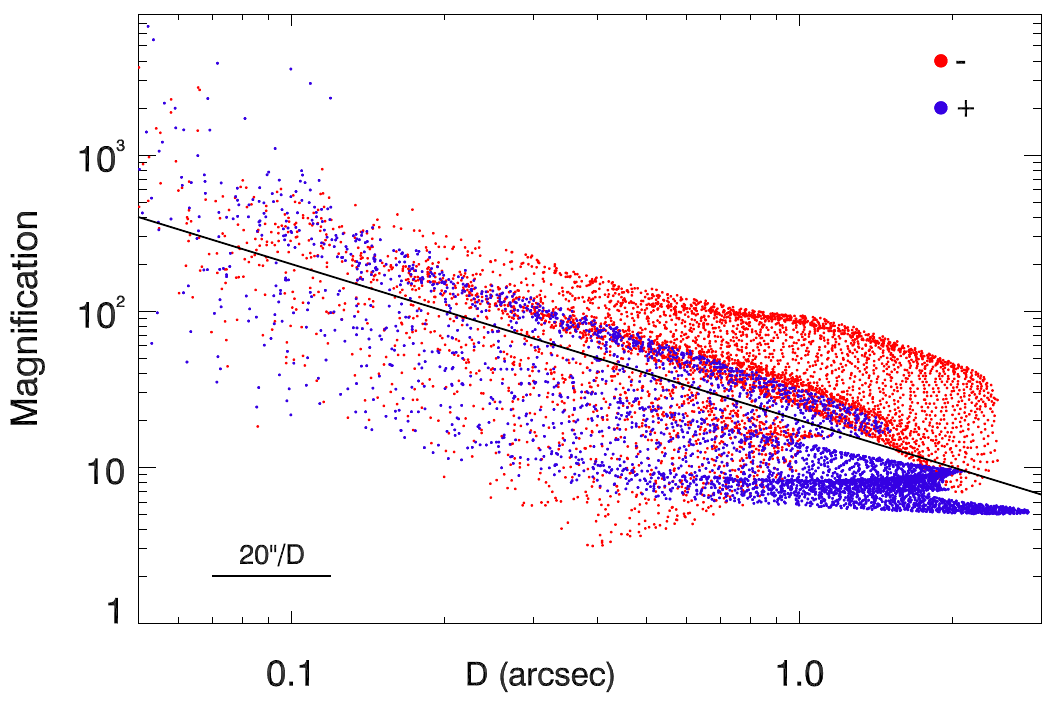}   
      \caption{Magnification versus distance for all pixels in the tail of the Dragon as a function of their distance, $D$m, to the CC. Red points are for pixels with negative parity while blue dots are for pixels with positive parity. The solid line shows the model $\mu=A/D(")$ with $A=20"$. Most pixels follow similar laws but with different values of $A$ 
      }
         \label{Fig_Mu_vs_distance}
\end{figure}

To correctly interpret the transient events in the Dragon, one of the most relevant lensing properties is the magnification from the macrolens. We compute the magnification in all the pixels in the tail of Dragon and represent them in Fig.~\ref{Fig_Mu_vs_distance} as a function of the distance, $D$, to the CC. The head of the Dragon does not show transient events so we do not include them in this figure. They are farther away from the critical curve and have more modest magnification factors, mostly in the range 5--20.

\begin{figure} 
  \includegraphics[width=\linewidth]{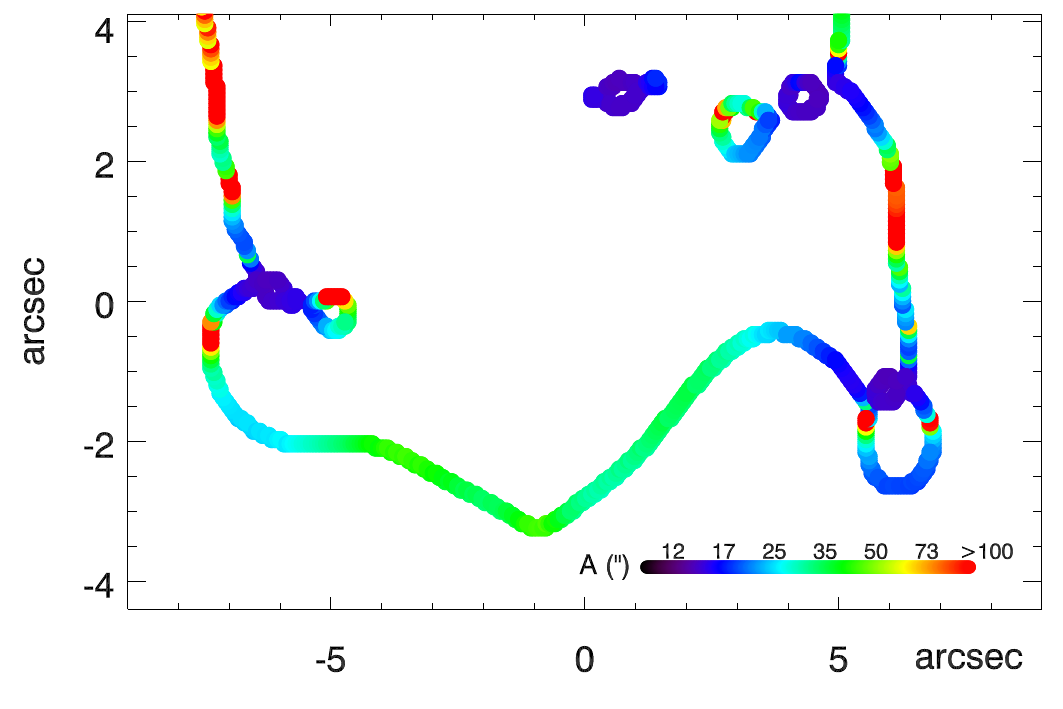}   
      \caption{Strength of the magnification along the critical curve in the Dragon arc region. The strength $A$ is the normalization of the relation $\mu=A/d$ where $d$ is the distance to the CC expressed in arcseconds. The normalization, $A$ is obtained after fitting the relation  $\mu=A/d$ in the direction of the gradient of the magnification from the \wslap model. 
         }
         \label{Fig_CC_Strength}
\end{figure}

The blue points in Fig.~\ref{Fig_Mu_vs_distance} are pixels in the portion of the Dragon with positive parity while red points are pixels in portions of the arc with negative parity. The parity is indicated in the bottom panel of Fig.~\ref{Fig_Dragon} with "+" and "-" symbols for the positive and negative parity respectively. As expected, the magnification in the regions of negative parity is more extreme (larger and smaller) than the magnification in the portions of the arc with positive parity. Also not surprisingly, we recover the canonical $\mu = A/D$ (represented by the black solid line for the case $A=20"$) although with significant scatter in the amplitude, $A$. This scatter is perhaps one of the main parameters describing the magnification properties along the Dragon arc. We fit for this parameter along all positions in the CC by first determining the direction of the CC at each position and fitting for $A=\mu\times D$ in a perpendicular direction. The result is shown in Fig.~\ref{Fig_CC_Strength} where we plot the value of $A$ for each point in the CC. Along the CC the strength factor, $A$, changes significantly. In particular, $A$ is smaller near member galaxies since the deflection field acquires more curvature near galaxies and the magnification is inversely proportional to the curvature of the deflection field. 
The events listed in \cite{Fudamoto25}, shown in the bottom panel of Fig.~\ref{Fig_Dragon}, are consistent with this picture since we observe a widening of the events towards the middle of the arc, where $A$ is largest. The width, $\Delta$, of the region where microlensing events are observed, can be approximated by $\Delta\approx 2d_{\rm crit}$, where $d_{\rm crit} = A\Sigma_{*}/\Sigma_{\rm crit}$ \citep{Diego2018}. The width of the distribution of events around the CC can be used also to test models of wave DM as shown in \citep{Broadhurst2025}, and specifically for the Dragon arc. In this case, the width is closely related to the mass of the ultra-light Boson. Future events in the Dragon will provide a clearer picture of the width, $\Delta$, in the distribution of events around the CC, allowing us to constrain $A$ directly from $\Delta$ and set limits on the boson mass in the case of wave DM models.

The time delay along the arc is another interesting lensing feature that needs to be studied in detail. Some transient events may be the result of intrinsic flux variations in small sources in the arc. For instance SNe or intrinsically variable stars, such as Luminous Blue Variables (LBV), several of which are likely to be present in the sample of \cite{Fudamoto25}. Of special interest are Cepheid stars. As shown in \cite{Diego2025}, Cepheids (especially the long period ones) can be observed in the Dragon with JWST in relatively short exposures. Their use as standard candles for cosmological studies (for instance to infer the value of $H_0$) is questionable based on the results shown in Fig.~\ref{Fig_CC_Strength}, since the macromodel magnification can be undetermined by a factor larger than the precision required to derive cosmological parameters. However, Cepheids can be used to infer that unknown magnification and refine the lens models incorporating the new magnification information as additional constraints in the lens model. Long period Cepheids are exceedingly rare but very luminous. If one of these Cepheids is observed in the Dragon it will likely appear multiple times, each one with a delay with respect to the others. Instead of using them to derive $H_0$ from their distance modulus (which is degenerate with the unknown magnification), one could use the time delay between repeating Cepheids to infer $H_0$, since the time delay is inversely proportional to $H_0$. Long period Cepheids ($P>10$ days) are particularly interesting since they can be matched more easily (there should be very few long period Cepheids in the arc), and they can be observed even at larger separations from the CC (i.e., with more modest but more likely magnifications factors).

\begin{figure*} 
  \includegraphics[width=\linewidth]{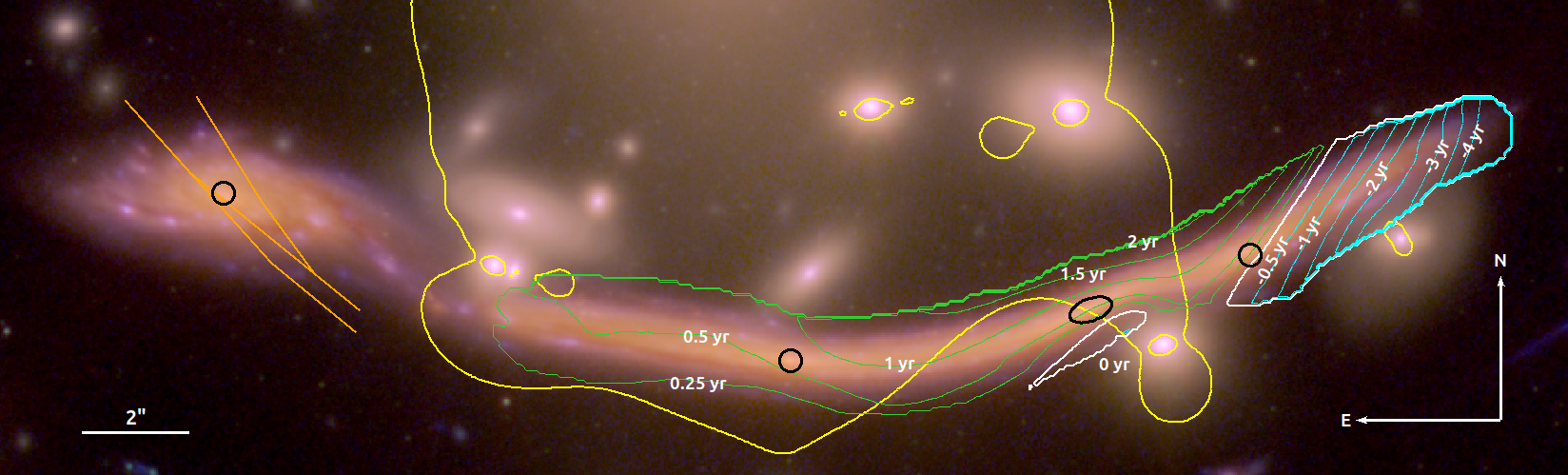}   
      \caption{Time delay along the Dragon arc. Contours show the isochrones of arrival time in the tail of the Dragon with respect to the corresponding position on the head of the Dragon. Images in the portion of the tail with cyan contours arrive before the images in the head while images in the region with green contours arrive after. The white contour marks images in the tail that arrive at the same time as in the head. The critical curve (yellow) and the nucleus (black circle) are included for reference. The orange lines crossing the head of the Dragon represent the caustics in the source plane remapped into the image plane. 
         }
         \label{Fig_TimeDelay}
\end{figure*}

We compute the time delay from the fourth model between every pixel in the tail of the Dragon and the corresponding pixel that maps into the head of the Dragon. The result is shown in Fig.\ref{Fig_TimeDelay} where we show the time delay as contours with intervals of 0.25 years or 0.5 years. For convenience we added the CC from the fourth model (yellow line) and marked with black circles or ellipses the position of the multiple images of the center of the galaxy. We also include the caustics remapped into the head of the Dragon (orange lines) to show the portion of the galaxy that is multiply imaged. Approximately half the galaxy has multiple images. From the figure we see how the west side of the arc arrives first, with the very west tip arriving $\approx 4$ years before the photons from the corresponding sibling area in the head of the Dragon to the east. When considering only the nucleus of the galaxy, photons from the west counterimage arrive almost simultaneously (white contours) to photons from the nucleus in the head of the Dragon. The other two images arrive also almost simultaneously but $\approx 0.7$ years after. More precisely, the lens model predicts a very small difference of 1.15 days between the arrival time of the two images of the nucleus in the central portion of the tail of the Dragon. 

In the new JWST MAGNIF observations (obtained in August 2024) we are able to find new transients (Fudamoto, private communication) that follow the distribution shown in the bottom panel of Fig.~\ref{Fig_Dragon}. One of the new events is located in a counterimage of the nucleus near the middle of the arc (second black circle from the left in Fig.\ref{Fig_TimeDelay} or knot a.2 in the middle panel of Fig.~\ref{Fig_Dragon}). This transient is unusual given its distance to the CC and the fact that near the nucleus we expect a large density of stars, resulting in very frequent microlensing events but with very small, and undetectable, flux changes. This is a variation of the more is less effect discussed in \cite{Welch2022}, and similar to the microlensing effect of distant globular clusters in \cite{Dai2021}. This event could be a different type of transient, for instance a type Ia SN that are more commonly found in evolved portions of galaxies. Another interesting alternative to microlensing is that we are witnessing a tidal disruption event at the super massive black hole (SMBH)  suspected to be lurking in the center of the galaxy \citep{Karmen2025}. Both in this case and the SN case, a reappearance of the same event should be observed in the other counterimages of the galaxy nucleus. From the time delay map, this counterimage is the last to arrive so any intrinsically variable source in this counterimage must have occurred in the other counterimages $\approx 0.7$ years prior to the JWST observations. 
No transient event is found in the other nearly simultaneous image (black ellipse in Fig.\ref{Fig_TimeDelay} or knot a.4 in the middle panel of Fig.~\ref{Fig_Dragon}) but this could be due to the fact that a.4 is a merging pair of images that misses part of the nucleus, which could be the part that is hosting the transient. Alternatively, the transient may have been too short lived and missed by the relatively short duration exposures of JWST.

Finally, regarding the possibility of observing Cepheids in the Dragon, more than 50 Cepheids are reported in a portion of NGC 4535 (at $16\pm1.9$ Mpc) \citep{Macri1999,Spetsieri2018}, a grand design spiral galaxy resembling the Dragon galaxy. Extrapolating to the entire galaxy we expect over 200 Cepheids in NGC 4535 brighter than absolute magnitude $M_V \approx -4$.  In nearby objects  the number of known Cepheids grows exponentially, with nearby galaxies such as the LMC harboring as many as 1333 known Cepheids \citep{Udalski1999}. Among the brightest, \cite{Persson2004} reports 13 Cepheids brighter than apparent magnitude 11.5 (J-band Vega) corresponding to absolute magnitude -7. Only the brightest Cepheids can be detected in the Dragon arc without the extra boost provided during a microlensing event. We explore the possibility of detecting Cepheids in the Dragon in the next section. 

\section{Discussion}\label{sect_discussion}

A370 is one of the best studied clusters in terms of the number of lensing constraints. The accurate mass and gravitational potential models allow us to use the cluster as a large-scale laboratory for gravitational studies and test models of dark matter or alternatives to GR. In addition, the Dragon arc offers unique opportunities to study a galaxy at cosmological distances with unprecedented detail, down to the level of individual stars in this galaxy. Below we explore these two applications of the lens model derived in the previous section in more detail. 

\subsection{Testing alternative DM models}
Given the relatively short distance from the two radial arcs to the center of the potential, it is useful to discuss our findings in the context of alternative models to $\Lambda$CDM. Lensing has already been used to challenge MOND models \citep{Sanders1999,Zhao2006,Natarajan2008}, most notably by the bullet cluster \citep{Clowe2004}, although efforts have been made to explain the bullet cluster within the MOND framework \citep{Aungus2006}. 
Usually, the role of DM is to account for missing mass at larger radii. Typically, one finds that isothermal models for the DM ($\rho_{3D}(r) \propto r^{-2})$ provide a good fit to rotational curves and lensing observations. This translates in DM contributing to the integrated mass as $M(<r) \propto r$ at large radii. MOND simply transforms this  $M(<r) \propto r$ from DM models into a change in the potential. In MOND models, the gravitational force is modified by a factor $f(r/r_o)$, where $f(r/r_o)\approx1$ when $r<<r_o$ and  $f(r/r_o)\propto r$   when $r>>r_o$ \citep{Milgrom1983,Sanders2002,Famaey2005}. This  $f(r/r_o)\propto r$ behavior at large radii in the potential is similar to the $M(<r) \propto r$ from DM (in an isothermal model). In MOND, when the condition  $r>>r_o$ is satisfied the gravitational acceleration is small, $a_o\approx 1.2\times10^{-10}$ m s$^{-2}$, \citep{Famaey2012}. It is at this acceleration where the presence of DM, or the departure from GR in MOND, can be checked with observations. In both cases, DM or MOND, at larger accelerations (or smaller radii) the potential is usually dominated by baryons (in the case of DM), or it behaves as the usual Newtonian potential (in the case of MOND). 
We can compute the gravitational acceleration at $r=23$ kpc (the aperture considered for the compact component around the two BCGs) from the estimated stellar mass in the two BCGs, $M_{\rm BCG}\approx 4.5\times10^{11}$\Msun\, and find $a=1.19\times10^{-10} {\rm m}\,{\rm s}^{-2}$. This is comparable to the characteristic scale of MOND, at which the effect of DM usually manifests. 

\begin{figure} 
  \includegraphics[width=\linewidth]{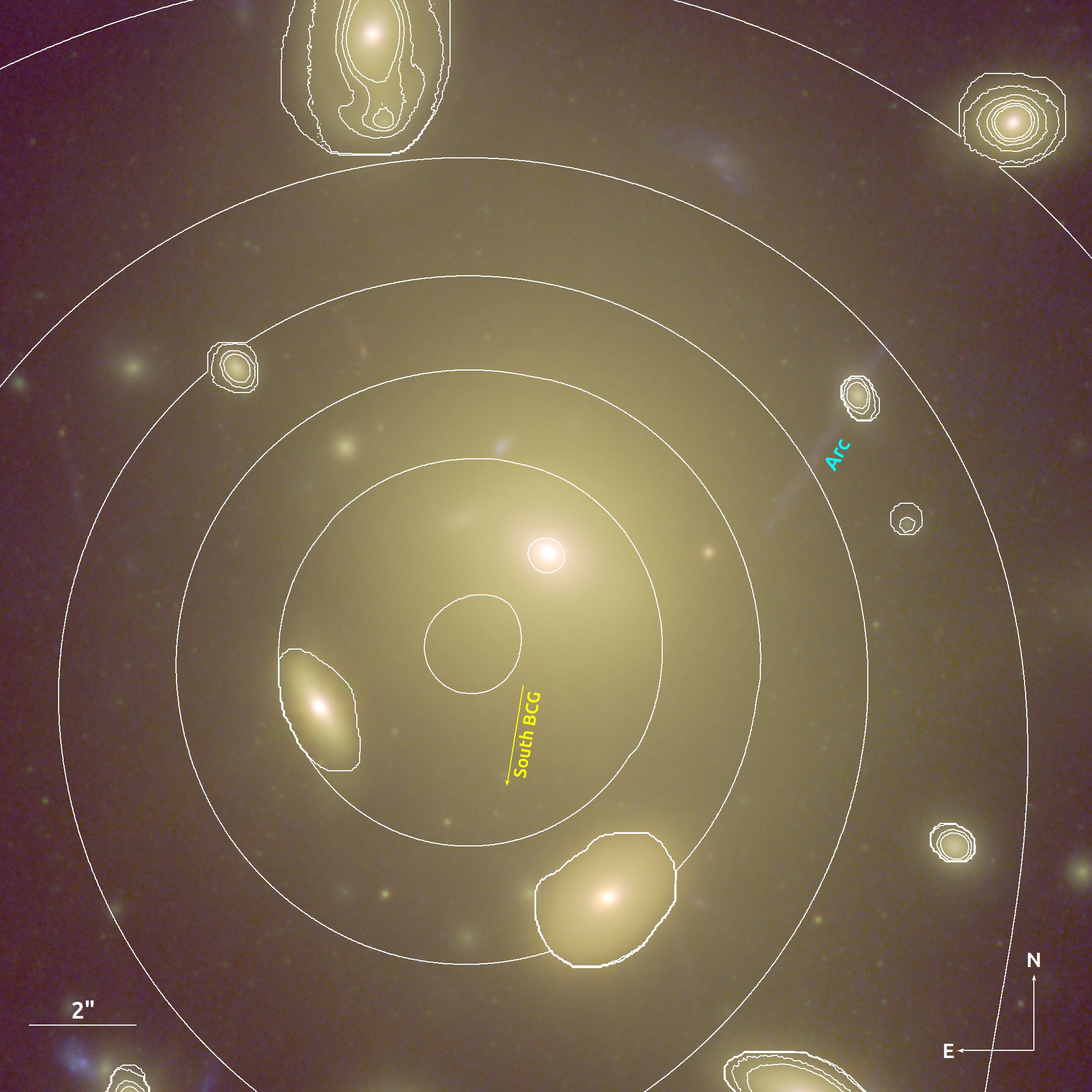}   
      \caption{Distribution of mass around the north BCG. The white contours show the total projected mass for model C around the north BCG. The peak of the soft component is displaced $\approx 2"=11.32$ kpc) from the center of the BCG. The yellow arrow shows the direction to the south BCG. }
         \label{Fig_NorthBCG_ModelC}
\end{figure}

If we focus on the best model for the north BCG (model C), we find that up to the position of the radial arc, the stellar mass alone is sufficient to explain that arc with no DM required within that galaxy to produce the required deflection field. This is a surprising result as it would imply a BCG galaxy with no DM. Taking a closer look at model C near the north BCG, the lens model places the peak of the smooth mass $\approx 11.32$ kpc to the southeast of the BCG center (see Fig.~\ref{Fig_NorthBCG_ModelC}). In this case the DM mass peak would be offset from the BCG and could explain the apparent deficit of DM at the BCG position. Interestingly, the direction of this offset is partially aligned with the alleged direction of the collision between the two groups (assumed to be in the direction of the south BCG and marked with a yellow arrow in the figure). We can only speculate that the DM and stellar mass components have been partially decoupled as a consequence of the interaction with the south BCG with both  components sloshing around each other as they settle after the interaction. Interestingly, an offset DM halo is part of the parametric model of \cite{Lagattuta2019} (offset$\approx 50$ kpc toward south-east). Also, the peak of the DM distribution is directly offset in the free-form model of \cite{Ghosh2021} (offset$\approx 50$ kpc toward south-east). 
Offsets of this magnitude have been observed also between the X-ray peak and the BCG \citep{Cross2024}, but in this case the offset may be purely due to hydrodynamical effects of the X-ray plasma and not affecting the stellar (and DM) component. Using gravitational lensing, \cite{Harvey2017} finds also offsets in the range of 10 kpc in galaxy clusters. These types of offsets are expected in self-interacting DM (SIDM) models \citep{Kim2017}, where offsets of this magnitude are expected for cross sections $\sigma/m \approx 1$ gr cm$^2$ g$^{-1}$ \citep{Harvey2019}. The offset in this model is the consequence of the BCG wobbling around the minimum of the potential. For this mechanism to work, the powerful DM spike predicted in $\Lambda$CDM needs to be significantly diminished, something that is naturally predicted in SIDM models. Offsets are also often observed in $N$-body simulations, specially after gravitational interactions \citep[see for instance][]{Dattathri2025}.
An alternative mechanism for evaporating the central DM spike is scouring by a binary super massive black hole \citep{Postman2012} or gravitational wave recoil of a merged binary SMBH \citep{Nasim2021,Khonji2024}. In any case, the north BCG of A370 represents an interesting study case that deserves further attention both at the modeling level, but also at the data level (for instance by performing a kinematic study based on deep spectroscopic data, a study well beyond the scope of this paper). The JWST data does offer an intriguing difference between the north and south BCGs, with the ICL around the north BCG appearing more puffy than the ICL around the south BCG. Perhaps a consequence of a heating mechanism inside the north BCG that is not as efficient in the south.  A yet another alternative scenario where offsets between the DM distribution and BCG are expected is in wave DM where random walk from the soliton can result in displacements of several kpc with respect to the BCG \citep{Schive2020,Chowdhury2021,XinyuLi2021}. However, in this case displacements of O(1) kpc are expected, well below the offset in model C. Larger displacements would be possible, but only for uncomfortably smaller masses of the boson, $m_B<<10^{-22}$ eV, already in tension with several observations. Even though the north BCG is intriguing and deserves more attention in the future, we also note that, from an epistemological perspective, the validity of our results regarding the apparent lack of DM in the north BCG and the observed offset in model C are questionable. In particular because our alternative model C', which includes the correct amount of baryonic matter in the BCG, performs more poorly when reproducing the north radial arc than the less physically motivated model C. This possibly signals to a missing ingredient or lack of flexibility in our model near the north BCG. This deserves more careful attention in future work.

The situation is different in the south BCG. In this case, at $\approx 23$ kpc, the mass, and hence acceleration, from the best models needs to be 6.6 times larger than the one produced by the stellar mass alone. This challenges the MOND interpretation since at this acceleration ($\approx 7\times10^{-10}\,{\rm m}\,{\rm s}^{-2}$) we are well within the Newtonian regime of MOND and the baryonic component should be sufficient to reproduce the radial arc morphology. On the contrary, our findings favor the existence of $\approx 85\%$ unaccounted mass (that is DM) within the inner 23 kpc. 
An alternative way of making the same argument is to find the separation from the BCGs at which the  $f(r/r_o)$ term in MOND should start to dominate and mimic the DM behavior, that is, the point at which $a=GM/r^2\approx 10^{-10}\,{\rm m}\,{\rm s}^{-2}$. From our best lens model, this happens at $\approx 60$ kpc or $\approx 11".6$ from the south BCG, and well beyond the point from the radial arc that is closest to the south BCG. Hence, we can argue that the radial arc near the south BCG is within the region where the departure from Newtonian behavior in MOND  should not be relevant, yet we still find that in the best lens model there is a need for a factor 6.6 times more mass than the stellar mass within the distance to the radial arc. 

A possible solution for MOND in this case is to consider extensions of the model. For instance the value of $a_o$ scales with the potential, with $a_0\approx 10^{-10}\, {\rm m}\,{\rm s}^{-2}$ for galaxies but larger values, $A_o>a_o$,  for clusters. Attempts in this direction include Extended MOND \citep{Zhao2012}. 
We use Eq. (9) in \cite{Hodson2017}, $A_o=a_o exp(|\phi/\phi_o|)$, where $A_o$ is estimated from 12 clusters using X-ray data and at different radii, or potential strengths, and with the constraint that X-ray data must be consistent with a MOND-type of model. The scale potential is fixed to $|\phi_o|=2.25\times10^{12}\, {\rm m}^2\,{\rm s}^{-2}$, and we estimate $|\phi|=8.45\times10^{10}\, {\rm m}^2\,{\rm s}^{-2}<<|\phi_o|$ from the stellar mass up to 23 kpc. At this separation, $A_o\approx a_o$, and we are still in the same regime as in the discussion above. One needs to move farther away from the BCGs in order to get $|\phi|\approx |\phi_o|$ and allow for a significant departure of $A_o$ with respect to $a_o$. A similar conclusion is obtained if one instead considers the escape velocity at 23 kpc from the BCGs, $V_{\rm esc}\approx 400\, {\rm km}\,{s}^{-1}$, and compare it with Fig.\,(1) in \cite{Zhao2012}. In this case $A_o\approx 1.3a_o$, clearly below the needed $A_o > 6a_o$. 

From the discussion above relative to the north and south BCGs, we conclude that A370 is inconclusive with the north BCG preferring MOND-like or SIDM models and the south BCG favoring $\Lambda$CDM. A possible solution for this apparent contradiction is adding more constraints near the BCGs, which are relatively poorly constrained within the central 30 kpc. As shown in the appendix, alternative parametric models based on almost the same set of constraints cannot reproduce the radial arcs better than our model C for the north radial arc and model A for the south radial arc, so these two models are still the best models for these arcs.

\begin{figure} 
  \includegraphics[width=\linewidth]{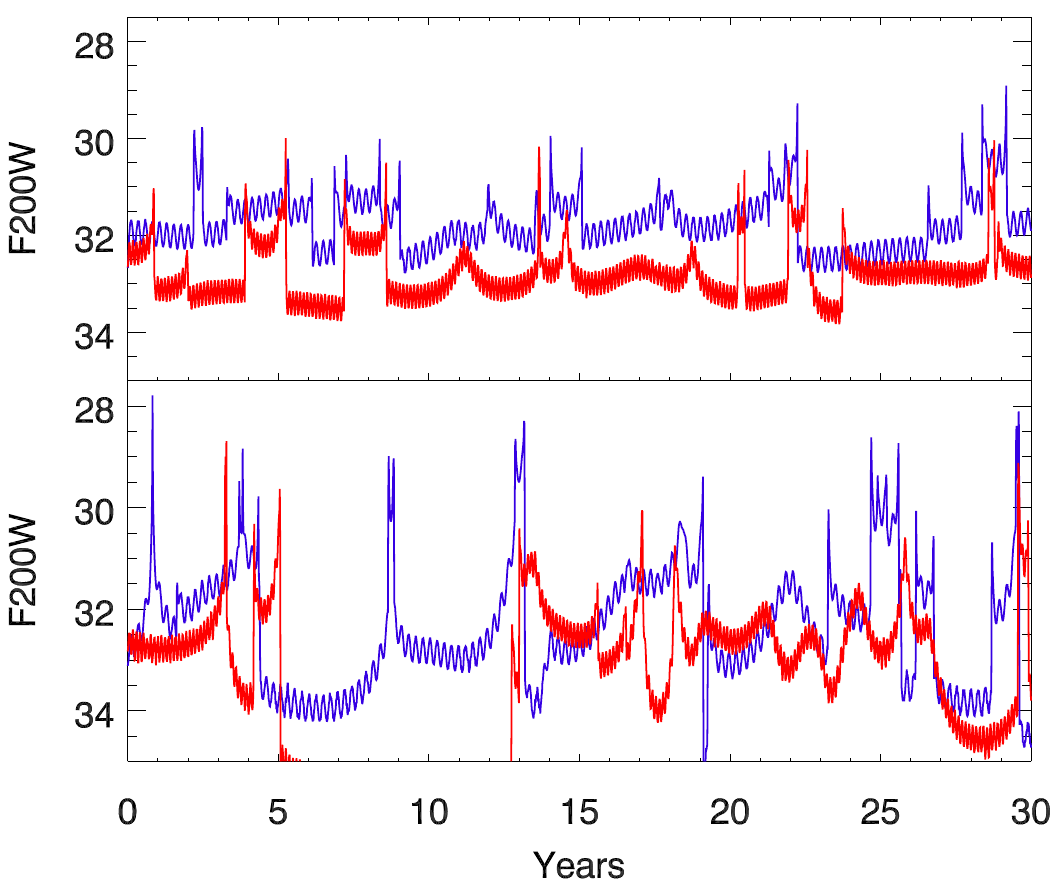}   
      \caption{Simulated light curve of two Cepheids under the effect of microlensing with periods 25 days (red) and 50 days (blue). The top panel shows the case of Cepheids moving across the web   
      of microcaustics in a portion of the Dragon with positive parity. The bottom panel shows the same Cepheids but on the portion of the Dragon with negative parity. In both cases, the macromodel magnification is $\mu=161$ and the observations is assumed to be carried out in the JWST's F200W filter (corresponding J-band in the rest frame for a Cepheid in the Dragon). For the blue curve in the bottom panel ($P=25$ days and negative parity), the Cepheid is brighter than  30.5 mag $8\%$ of the time while the red curve ($P=25$ days and negative parity) is above this limit only $1\%$ of the time. } 
         \label{Fig_Cepheid}
\end{figure}

\subsection{Cepheids in the Dragon}
\cite{Diego2025}  suggested that long-period classic Cepheids ($P>30$~days) could be observed in the Dragon arc. We study this possibility in more detail here by considering also the effect of microlenses. We consider a small region in the source plane with macromodel magnification $\mu=161$. This is representative of the regions near the critical curve where we expect to see most of the events, which are regions in the range of macromodel magnifications $100<\mu<300$. For smaller macromodel magnifications, the brightest Cepheids will remain undetected in deep exposures with JWST even after accounting for the extra boost in magnification provided by microlenses. For larger magnification factors, the area in the source plane is so small that the probability of finding a long period Cepheid is too low.  As shown earlier, a typical value for the strength of the CC around the Dragon is $A\approx 50"$. Macromodel magnification in the range $100<\mu<300$ would then correspond to separations from the CC in the range $0\farcs167<d<0\farcs5$

To simulate the magnified light curve of a Cepheid under the influence of microlenses we assume a surface mass density of microlenses of $\Sigma_*=40$ \Msun\,pc$^{-2}$. This is consistent with earlier estimates of $\Sigma_*$ \citep{Li2024}. The exact value of $\Sigma_*$ plays a minor role in the following discussion. Lastly, we simulate the light curve of luminous Cepheids in the Dragon by adopting a simple sinusoidal model with period $P$ and amplitude 0.5 magnitudes difference between the maximum and minimum. This is a typical variation for many classic Cepheids \citep[see for instance][]{Ulaczyk2012}. The sinusoidal shape is also a good approximation (fundamental mode) although many Cepheids can exhibit more complicated patterns. Here we ignore overtone modes for the Cepheids since we simply want to illustrate the lensed light curve in a simple scenario. We also factor time dilation in the observer plane (where caustics form) which effectively stretches the period by a factor $(1+z).$ 
Given the intrinsic (i.e., before time dilation) period of a Cepheid, we estimate the absolute magnitude in a given band following \cite{Storm2011},  $M_J = -3.18[{\rm log}_{10}(P)-1] -5.22$,   
where $P$ is in days and $M_J$ is the absolute magnitude (Vega system) in the J-band. This band, when the Cepheid is in the Dragon redshifts into $\sim 2$ micron, so we consider the filter F200W from JWST as the observation band of these Cepheids with magnitude $M_J$ in the rest-frame J-band. We transform Vega to AB magnitudes following \cite{MaizApellaniz2007} and apply the band correction, $2.5{\log_{10}(1+z)}$ for $z=0.7251$. Then we correct for the distance modulus to this redshift, 43.12.

\begin{figure} 
  \includegraphics[width=\linewidth]{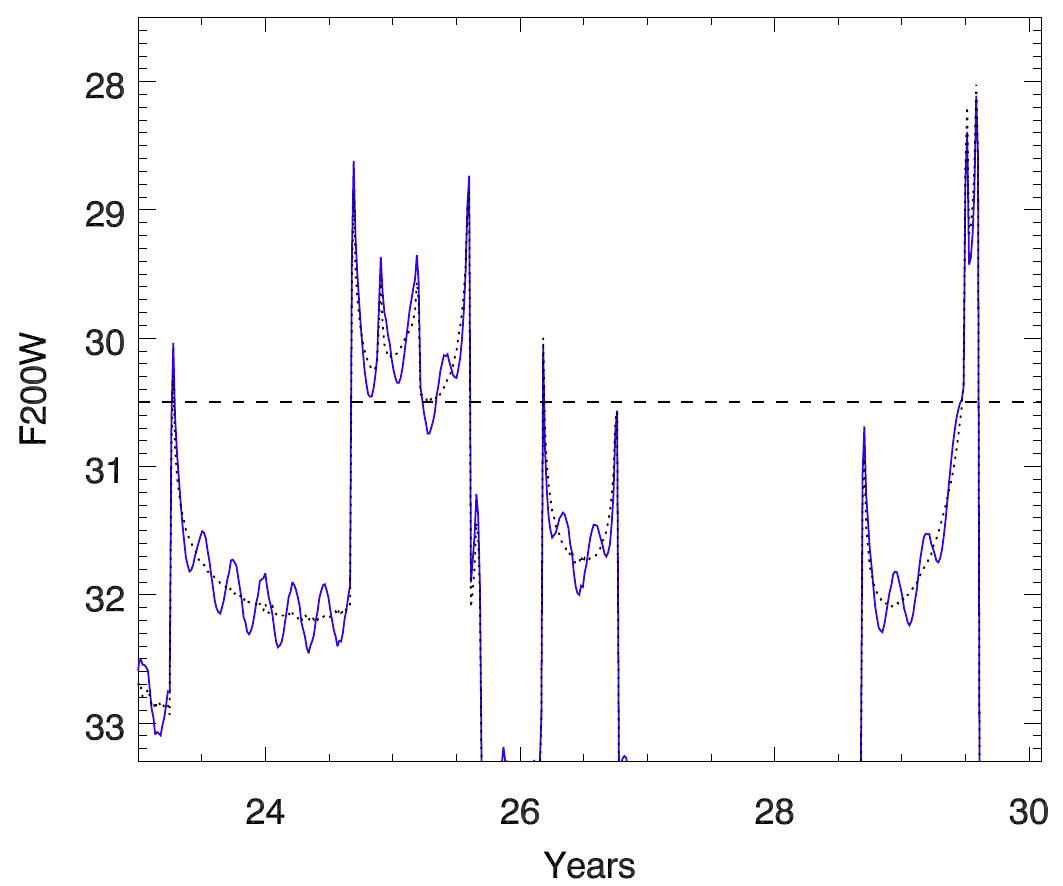}   
      \caption{Zoom in a portion of the light curve for the 50 day period Cepheid with negative parity in Fig.~\ref{Fig_Cepheid}. The dotted line shows the effect of magnification from the macromodel and microlenses. 
      The dashed line marks the detection limit in this filter with JWST in deep observations ($\approx 15$ hour exposure.)}
         \label{Fig_Cepheid_Zoom}
\end{figure}

Finally, the light curve of the Cepheids are corrected for magnification by subtracting the term $2.5{\rm log}_{10}(\mu_{\rm micro})$, where $\mu_{\rm micro}$ is the magnification after accounting for microlensing effects. We also need to assume a transverse relative velocity, $v_{\rm rel}$, between the Cepheids and the web of microcaustics. For this discussion we assume a transverse relative velocity $v_{\rm rel}=1000$ km s$^{-1}$ which is a typical velocity for the redshifts of A370 and the Dragon  \citep{Kayser1986,MiraldaEscude1991,Kelly2018}. With this velocity, and for the resolution of the microlensing simulation (2 nanoarcsec/pixel) the light curves are sampled every $\approx 3$ days.

The resulting light curves for two Cepheids with $P=25$ and $P=50$ days are shown in Fig.~\ref{Fig_Cepheid} as red and blue curves respectively. Since the Dragon has portions of the tail where the parity is positive, and portions where the parity is negative,  we consider both scenarios. As shown in the figure, the portion with negative parity produces brighter events. This is a well known property of images with negative parity, where brighter peaks need to compensate the smaller magnification factors which are not present in images with positive parity \citep{Diego2018}.

The Cepheid with $P=25$ days (red curve) remains undetected for a period of 30 years in observations reaching AB 30 in the portions of the Dragon with positive parity (top panel). In contrast, in the portion with negative parity (bottom panel), the same Cepheid makes 3 or 4 brief appearances brighter than this magnitude during the same period, but they are too short to reliably allow us to estimate the period, or even recognize the star as a Cepheid. Only Cepheids with periods $P>50$ days can be detected during enough epochs to allow us to measure their periods. 

Zooming in a portion of the light curve for the $P=50$ days Cepheid on the side with negative parity, we can see the pulsation of the Cepheid,  Fig.~\ref{Fig_Cepheid_Zoom}. In deep observations reaching $\sim 30.5$ AB in F200W, this Cepheid can be observed pulsating regularly $\approx 6$ times over 1 year. Brighter Cepheids can be observed more easily during fainter microlensing events or at smaller macromodel magnification factors (that is, farther away from the CC). The dotted line in the figure represents the effect of the magnification, and can be viewed also as the light curve of a similar star with the same magnitude, $M_J$, but without oscillations. As shown in the figure, at around 25 years there are two small fluctuations in the magnification from overlapping microcaustics that perturb the periodic oscillation of the Cepheid. A more smooth behavior can be appreciated on the microlensing event at around 29 years, although in this particular case, the Cepheid remains too faint to be observed for most of the duration of the microlensing event. Overlapping microcaustics represent a challenge to identify and model lensed Cepheids since they can break the periodicity in the light curve, or produce false positives with stars that are not truly Cepheids. Nevertheless, the possibility of observing long period Cepheids in the Dragon is real and should be considered in future studies of the transients in this interesting arc. In this regard, a possible complication for Abell 370 is its low latitude, with two relatively narrow viewing windows of $\approx 1.5$ month with JWST: August and December, impacted by meteoroid avoidance restrictions. This will hamper the monitoring efforts carried out with JWST for long-period Cepheid. However, upcoming instruments, such as MICADO at ELT ($\sim$ 2030), will have more flexibility in the monitoring throughout the fall semester, with a sensitivity and resolution surpassing those of NIRCam.

\section{Conclusions}\label{sect_conclusions}
We revisit the galaxy cluster A370 by combining the JWST data from the CANUCS, GO-3538 and MAGNIF program. We derive a new lens model for A370 based on previous constraints from the literature, from which we redefine three systems based on the new MAGNIF-processed data and expand the set of constraints with two additional systems (one of them a known triply imaged SN). The lens model shows high consistency with earlier lens models. The new data discovers new member candidates of a  previously identified galaxy overdensity at $z\approx 1$ behind the cluster. We focus on the distribution of mass around the two BCGs, using two radial arcs from the same background galaxy near the BCGs to select the best lens models. From the best models we derive the fraction of DM  within the BCGs and find that the north BCG is consistent with containing no DM within 23 kpc from the center of the BCG. In contrast, the best model for the south BCG contains 6.6 times more mass than the stellar mass. We discuss these findings in the context of MOND and conclude that the north BCG favors MOND-like models while the south BCG favors $\Lambda$CDM. This apparent contradiction may be due to the scarcity of lensing constraints in the inner 30 kpc around the BCGs. A similar contradiction is found when considering SIDM models, where the offset between the DM peak and the north BCG is expected in SIDM models but such offset is not observed in the south BCG. The dichotomy in our results can be resolved with additional data that is able to better constrain the inner region. For instance, kinematic data from spectroscopic observations of the BCGs can provide yet another constraint on the minimum of the potential, but that study is beyond the scope of this paper. 

We also use the lens model to study in detail the lensing properties of the Dragon arc, a caustic crossing galaxy at $z=0.725$ where tens of microlensing events from supergiant stars at this redshift have been claimed from recent JWST data. We compute the strength, $A$, of the CC along the Dragon and find that it varies a factor $\approx 2$ between different portions of the arc. This variation can be estimated directly with future data by measuring the width of the region containing microlensing events around the CC. We also compute the arrival time of different portions of the arc finding differences up to $\approx 4$ years between different portions of the arc. Finally we study the possibility of observing bright Cepheids in the arc and the role played by microlensing. We show how detection of Cepheids with periods larger than $\sim 50$ days is possible but their identification as a Cepheid is challenging due to microlensing. 

\begin{acknowledgements}
J.M.D. acknowledges the support of projects PID2022-138896NB-C51 (MCIU/AEI/MINECO/FEDER, UE) Ministerio de Ciencia, Investigaci\'on y Universidades and SA101P24 (Junta de Castilla y Le\'on).
ML acknowledges CNRS and CNES for support. 
RAW acknowledges support from NASA JWST Interdisciplinary Scientist grants
NAG5-12460, NNX14AN10G and 80NSSC18K0200 from GSFC. We thank the anonymous referee for  constructive and useful comments.

This work is based on observations made with the NASA/ESA/CSA \textit{James Webb} Space Telescope. The data were obtained from the Mikulski Archive for Space Telescopes at the Space Telescope Science Institute, which is operated by the Association of Universities for Research in Astronomy, Inc., under NASA contract NAS 5-03127 for JWST. 
These observations are associated with JWST programs \#1208, \#2883 and \#3538. Support for program \#2883 was provided by NASA through a grant from the Space Telescope Science Institute, which is operated by the Association of Universities for Re- search in Astronomy, Inc., under NASA contract NAS 5-03127.

\end{acknowledgements}

\bibliographystyle{aa} 
\bibliography{MyBiblio} 

\begin{appendix}

\section{\label{sc:model_copmparison}Comparison with other lens models}

\begin{figure} 
  \includegraphics[width=9cm]{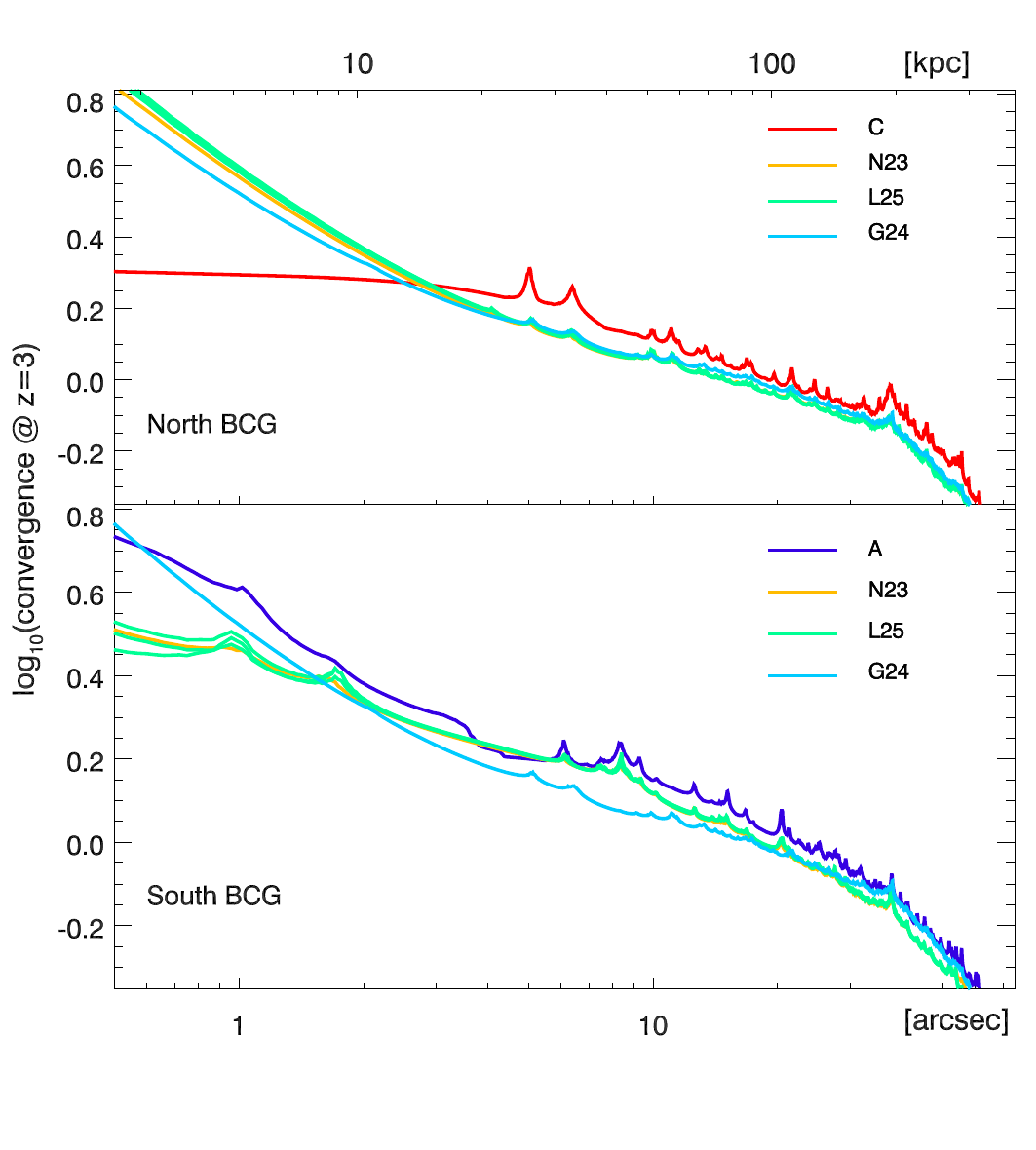}   
      \caption{The orange lines, N23, are for the parametric model in \cite{Niemiec2023}. Green lines, L25, are for the three models in \cite{Limousing2025}. Light-blue lines, G24, are for the model in \cite{Gledhill2024}. The red line in the top panel is for our model C, while the dark-blue line in the bottom panel is for our A model. 
         }
         \label{Fig_Profile_ALL}
\end{figure}

The \wslap solution presented in this work is a hybrid model that makes minimal assumptions about the distribution of mass. It is important to compare our model with previous parametric models that are based on a  different set of assumptions. For this comparison we consider the models from \cite{Niemiec2023} or N23, \cite{Gledhill2024} or G24, and \cite{Limousing2025}. For the latter, we consider the three reference models with the smallest RMS presented in that work. We refer to these three models as L25-1 (RMS=0".7), L25-2 (RMS=0".7), and  L25-3 (RMS=0".73). 
All models use very similar lensing constraints, and also similar to the ones used in this work. The parametric models are all derived using the code {\small LENSTOOL} \citep{Kneib1993,Jullo2007,Jullo2009}. 

\begin{figure} 
  \includegraphics[width=\linewidth]{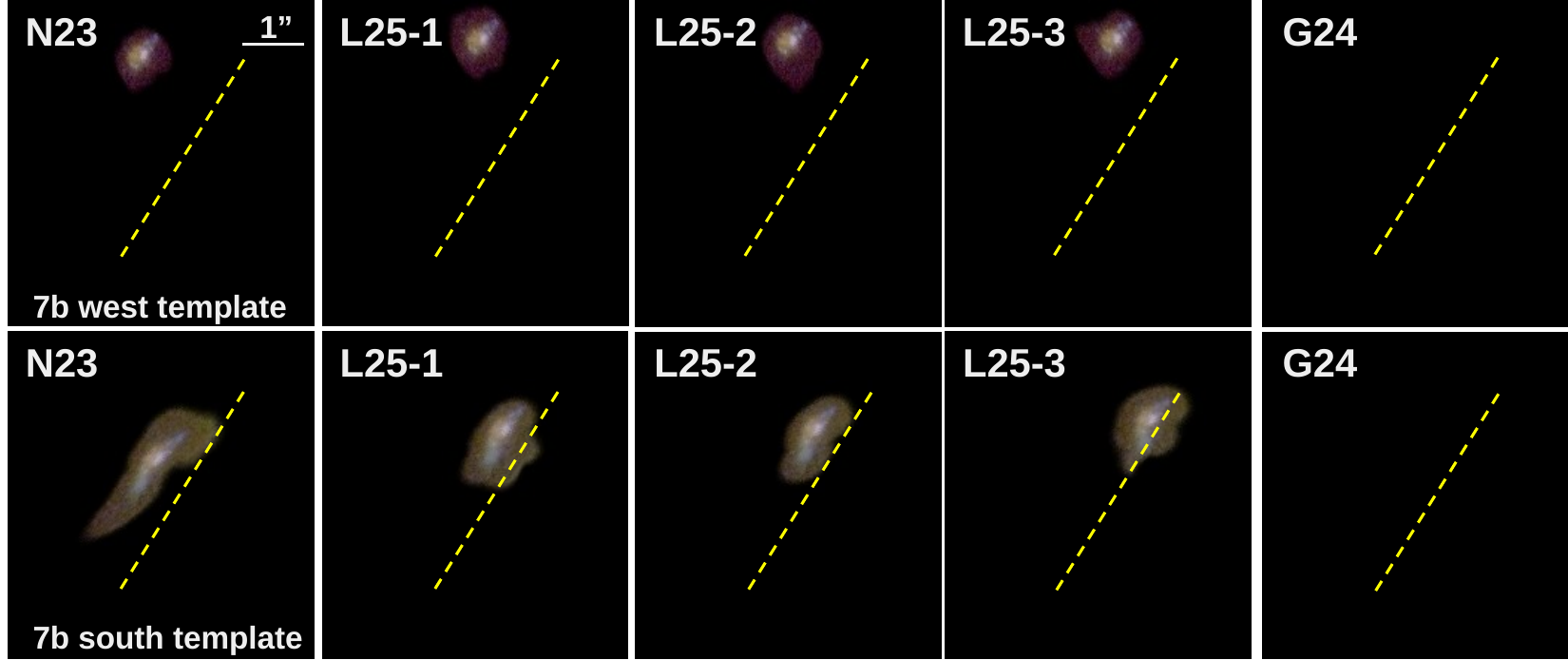}   
      \caption{Similar to Fig.~\ref{Fig_System7b} but for the parametric models. Predicted radial arc 7a near the North BCG for models N23, L25 and R24. The top row is the prediction using the west counterimage of system 7 as template. The bottom panel is the predicted arc when the counterimage in the south is used as template.  The dashed line marks the position of the observed arc.
         }
         \label{Fig_System7b_ALR}
\end{figure}

We compare the profiles around the north and south BCGs in Fig.~\ref{Fig_Profile_ALL}. For reference we include the profiles from our best models, C for the north BCG and A for the south BCG. For the north BCG all parametric model profiles are very similar. They predict a steeper profile in the inner 20 kpc than our best model C.  However, this steep profile cannot reproduce well the arc 7b near the north BCG as shown in Fig.~\ref{Fig_System7b_ALR}. None of the parametric models predict a merging arc near the observed arc (the G24 model predicts the counterimage outside the region shown). In contrast, our Model C (red curve in the top panel of Fig.~\ref{Fig_Profile_ALL}) is able to produce a merging pair of counterimages close to the observed position (1" or less) and with the right orientation (see Fig.~\ref{Fig_System7b}. The profile from model C is significantly shallower than the parametric models but between 20 and 30 kpc it contains more projected mass than the parametric models, compensating the mass deficit in the central region. Arc 7b favors this cored profile rather than the steep parametric ones. 

\begin{figure} 
  \includegraphics[width=\linewidth]{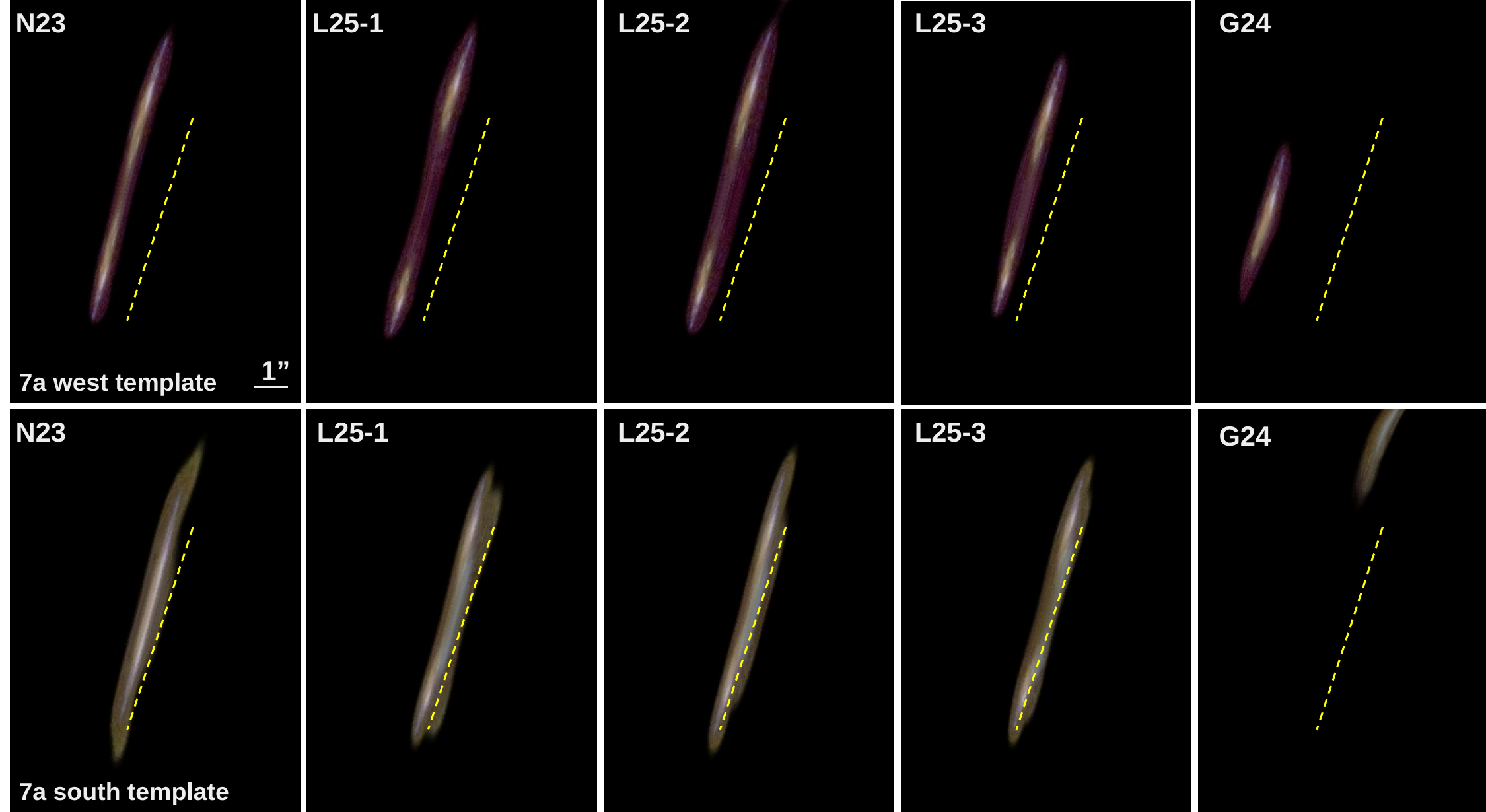} 
      \caption{Similar to Fig.~\ref{Fig_System7a} but for the parametric models. Predicted radial arc 7a near the South BCG for models N23, L25 and R24. The top row is the prediction using the west counterimage of system 7 as template. The bottom panel is the predicted arc when the counterimage in the south is used as template. The dashed line marks the position of the observed arc.
         }
         \label{Fig_System7a_ALR}
\end{figure}

The situation is different for the south BCG. The parametric profiles are in better agreement with the best hybrid model A. In the range 20--40 kpc, the parametric models from N23 and L25 are in good agreement with the hybrid model, while the model from G24 is slightly below. The situation reverses at large radii where G24 is in good agreement with the hybrid model in the range 200--300 kpc. The models from N23 and L25 reproduce the arc 7a reasonably well (see Fig.~\ref{Fig_System7a_ALR}),  especially when using the counterimage in the south as template. When the west counterimage is used instead, an offset of $\approx 1"$ is observed. This is larger than the offset in our model A for the same template image (see Fig.~\ref{Fig_System7a}). 

From the comparison above we conclude that among all models tested, the hybrid model C for the north BCG and the hybrid model A for the south BCG (and to a lesser degree  N23 and L25) reproduce arcs 7a and 7b best. Hence the discussion in section~\ref{sect_discussion} regarding the $\Lambda$CDM vs MOND is robust when considering other lens models.

\begin{figure*} 
  \includegraphics[width=\linewidth]{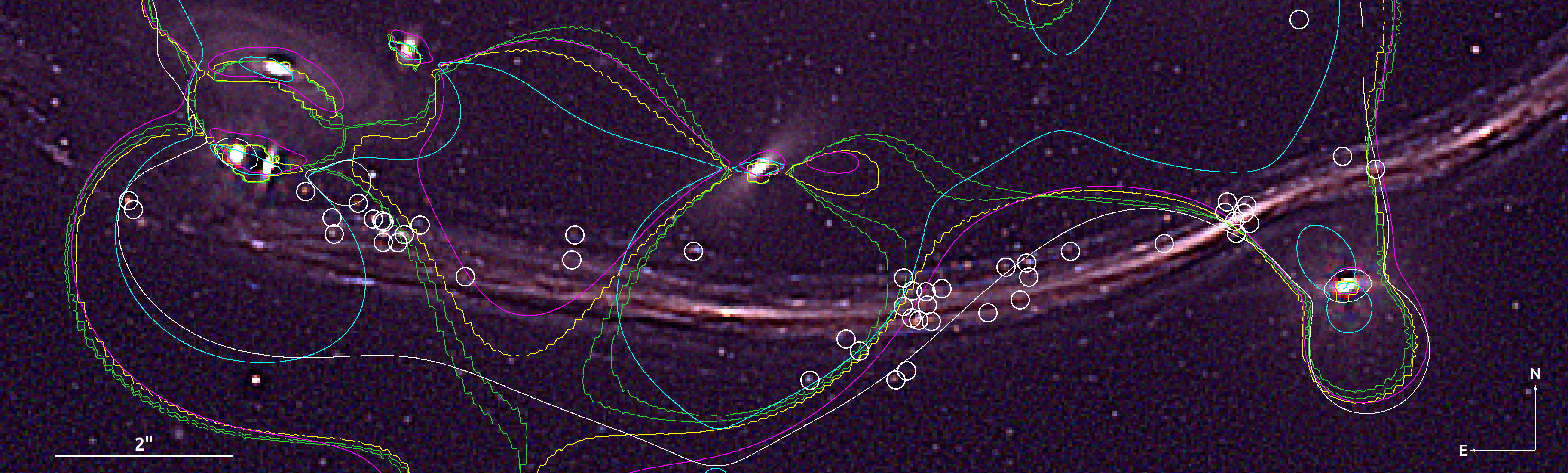}   
      \caption{As in the bottom panel of Fig.~\ref{Fig_Dragon} but showing the CC for the parametric models of \cite{Niemiec2023} (magenta), \cite{Gledhill2024} (cyan) and the three models in \cite{Limousing2025} (green and yellow). The yellow curve is the model L25-2 in \cite{Limousing2025}. For comparison, the same hybrid model from Fig.~\ref{Fig_Dragon} is shown as a white line.  
         }
         \label{Fig_Dragon_ARM}
\end{figure*}

Another interesting feature to compare is the CC around the Dragon. We compare the CCs from all parametric models in Fig.~\ref{Fig_Dragon_ARM}. The CC from N23 is shown as a white line. The three models from N25 are shown as green (L25-1 and L25-3) and yellow lines (L25-2). The model from G24 is shown in cyan. 

Among these models, the hybrid model,  N23 and L25-2 are more consistent with the distribution of transients in \cite{Fudamoto25} shown as circles. However, given the fact that microlensing events in the portion of the Dragon with negative parity can be magnified more, we expect more events on these portions of the arc. In this sense the hybrid model reproduces better this predicted property of the microlensing transients

The more convoluted structure of models in G24 (cyan), L25-1 and L25-3 (green) is also observed in some of the \wslap models reflecting the high uncertainty in the exact location of the CC in this portion of the lens plane, and the sensitivity to the lens model around the Dragon to small changes in the distribution of mass, especially from the massive member galaxy near the center of the figure. Additional detections of transients in the Dragon with future data will allow us to increase the number density of transients, map more precisely the location of the CC, and use that information to better constrain the distribution of mass around the Dragon. 

\end{appendix}

\end{document}